# Magnetic Anisotropy in Two-dimensional van der Waals Magnetic Materials and Their Heterostructures: Importance, Mechanisms, and Opportunities

Yusheng Hou[1], and Ruqian Wu[2,*]

[1] Guangdong Provincial Key Laboratory of Magnetoelectric Physics and Devices, Center for Neutron Science and Technology, School of Physics, Sun Yat-Sen University, Guangzhou, 510275, China

[2] Department of Physics and Astronomy, University of California, Irvine, CA 92697-4575, USA

**Abstract**

Two-dimensional (2D) magnetism in atomically thin van der Waals (vdW) monolayers and heterostructures has attracted significant attention due to its promising potential for next-generation spintronic and quantum technologies. A key factor in stabilizing long-range magnetic order in these systems is magnetic anisotropy, which plays a crucial role in overcoming the limitations imposed by the Mermin-Wagner theorem. This review provides a comprehensive theoretical and experimental overview of the importance of magnetic anisotropy in enabling intrinsic 2D magnetism and shaping the electronic, magnetic, and topological properties of 2D vdW materials. We begin by summarizing the fundamental mechanisms that determine magnetic anisotropy, emphasizing the contributions from strong ligand spin-orbit coupling of ligand atoms and unquenched orbital magnetic moments. We then examine a range of material engineering approaches, including alloying, doping, electrostatic gating, strain, and pressure, that have been employed to effectively tune magnetic anisotropy in these materials. Finally, we discuss open challenges and promising future directions in this rapidly advancing field. By presenting a broad perspective on the role of magnetic anisotropy in 2D magnetism, this review aims to stimulate ongoing efforts and new ideas toward the realization of robust, room-temperature applications based on 2D vdW magnetic materials and their heterostructures.

* Corresponding authors: wur@uci.edu

# 1. Introduction

Magnetic anisotropy is a fundamental and essential physical property of magnetic materials, describing the tendency of magnetizations to preferentially align along specific crystallographic directions. Microscopically, magnetic anisotropy can be broadly categorized into three primary components: (i) magnetocrystalline anisotropy (MCA), (ii) magnetic shape anisotropy (MSA), and (iii) exchange anisotropy. MCA, the most prevalent form of magnetic anisotropy, arises from the interaction between magnetic ions and their surrounding crystal field via spin-orbit coupling (SOC). This interaction introduces directional preferences for magnetizations that depend on the material's symmetry and electronic structure. In contrast, MSA originates from long-range dipole–dipole interactions among magnetic moments. A defining feature of MSA is its strong dependence on the geometry of magnetic materials.[1] While MSA is typically negligible in three-dimensional (3D) bulk magnets due to its relatively weak strength compared to MCA, it becomes increasingly significant in low-dimensional systems—particularly in magnetic thin films and quasi-two-dimensional materials.[2] Considering that two-dimensional (2D) van der Waals (vdW) magnetic materials are atomically thin, MSA is expected to play a crucial role in determining their magnetic properties. Exchange anisotropy, in contrast, arises from the interplay between SOC and the intrinsic crystal symmetry of magnetic materials. In general, lower crystal symmetries combined with strong SOC enhance the likelihood of exchange anisotropy, which in turn modifies the preferred directions of magnetizations and influences the overall magnetic ordering. This form of anisotropy is particularly significant in systems where interfacial effects or competing magnetic interactions are present. A clear understanding of the distinct contributions from all types of magnetic anisotropy—including exchange anisotropy—is essential for effectively tuning the magnetic properties of materials, especially in the context of spintronic applications and low-dimensional magnetic systems.

As research into 2D magnetism continues to progress, a growing consensus has emerged around the pivotal role of magnetic anisotropy in stabilizing long-range magnetic orders and enabling the emergence of complex magnetic phases in 2D magnetic materials. According to the well-known Mermin-Wagner theorem,[3] long-range ferromagnetic (FM) or antiferromagnetic (AFM) orders are forbidden at finite temperatures in isotropic Heisenberg spin systems due to the divergence of thermal spin fluctuations in low dimensions. This theoretical limitation, however, assumes the absence of magnetic anisotropy, which breaks continuous spin rotational symmetry and opens an energy gap in the spin-wave spectrum, thereby suppressing long-wavelength fluctuations that would otherwise destabilize magnetic orders. Experimental findings over the past several decades have clearly demonstrated that long-range magnetic ordering can, in fact, persist in 2D systems when magnetic anisotropy is present. Early studies in the 1960s revealed long-range FM order in ultrathin NiFe films as thin as a few atomic layers.[4] More recently, the isolation of atomically thin vdW magnetic materials has reinvigorated interest in this field. Landmark discoveries include the observation of long-range FM order in $CrI_3$ monolayers (MLs) in 2017,[5] and 2D long-range AFM orders in $FePS_3$ MLs in 2016.[6, 7] These findings not only suggest the crucial role of magnetic anisotropy in overcoming the constraints of the Mermin-Wagner theorem but also underscore its importance in guiding the design of 2D magnetic materials for applications in spintronics, quantum information, and topological devices.

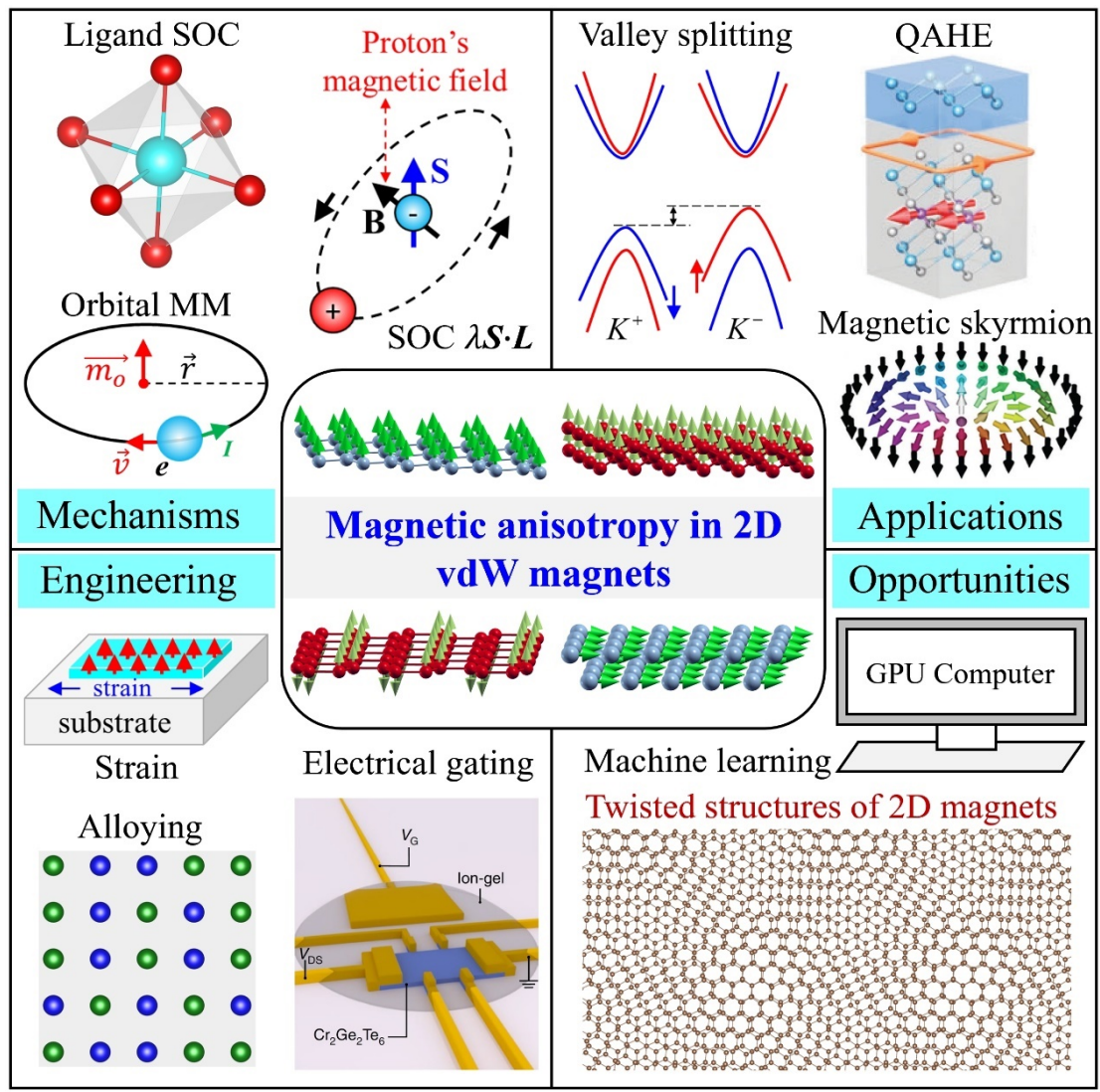

*Figure 1. A summary of issues discussed in this review. SOC: spin-orbit coupling; MM: magnetic moment; QAHE: quantum anomalous Hall effect. Sketches of QAHE,[8] magnetic skyrmion[9] and electrical gating[10] are reproduced with permission from*

*Refs.[9-11]. Ref.[8]: Copyright 2023, The Author(s), Ref.[9]: Copyright 2024, Wiley-VCH GmbH. Ref.[10]: Copyright 2020, The Author(s), under exclusive licence to Springer Nature Limited.*

Within the framework of linear spin wave theory, it is understood that in 2D magnetic materials with continuous spin rotational symmetry, the gapless spin-wave excitations lead to a divergence in magnetization at finite temperatures, which destabilizes the long-range magnetic order. To circumvent this issue, such a divergence has to be eliminated. It has been demonstrated that even a small out-of-plane magnetic anisotropy, which opens a finite spin-wave gap, can suppress the divergence and thereby stabilize long-range magnetic order in 2D magnetic systems.[12] Alternatively, long-range dipole-dipole interactions inherently break spin rotational invariance, allowing them to stabilize long-range magnetic orders in 2D magnetic materials as well.[2] Therefore, for long-range magnetism to persist in 2D magnetic materials, the continuous spin rotational symmetry must be broken, either through out-of-plane magnetic anisotropy or long range dipole-dipole interactions.

Due to the quenching of orbital magnetic moments in solids, MCA is typically very small, on the order of meV or sub-meV per magnetic atom. This weak magnitude makes it challenging to establish general principles governing both the magnitude and even the sign of magnetic anisotropy energy (MAE). As a result, developing a systematic approach to controlling magnetic anisotropy in most 2D magnetic materials remains difficult, despite its critical role in tailoring their functionalities and applications. Understanding magnetic anisotropy primarily relies on density functional theory (DFT) calculations, which have provided key insights into its microscopic origins. Detailed electronic structure analyses reveal intricate connections between the wavefunctions of states near the Fermi level, SOC matrix elements, and the bandwidths of specific magnetic materials. These findings highlight the fundamental role of electronic correlations and SOC-driven interactions in determining the anisotropic magnetic behavior of 2D magnetic materials.

By determining the preferred orientation of magnetic moments, magnetic anisotropy also plays a crucial role in shaping the electronic, magnetic, and topological properties of 2D vdW magnetic materials and their heterostructures. One notable consequence is the anisotropic magnetoresistance, which arises from changes in the Fermi surface induced by the rotation of magnetization direction.[13-15] When 2D ferromagnetism coexists with nontrivial topological band structures, magnetic anisotropy—particularly with out-of-plane magnetization—can enable the realization of exotic quantum states such as the quantum anomalous Hall effect (QAHE) and the axion insulator phase.[16]

Furthermore, when monolayer valley semiconductors are interfaced with 2D vdW magnetic materials, magnetization (either out-of-plane or in-plane) can induce valley splitting, thereby facilitating the practical implementation of valley Hall effects.[17] Magnetic anisotropy also interplays with other fundamental magnetic interactions, such as Heisenberg exchange and Dzyaloshinskii-Moriya (DM) interactions, which can stabilize topologically nontrivial spin textures, including magnetic skyrmions and bimerons.[9] Altogether, MLs or few-layer heterostructures of vdW magnetic materials provide a fertile platform for engineering anisotropy-driven functionalities.

From the perspectives of applications, 2D vdW magnets and their heterostructures offer tremendous potential in emerging fields such as spintronics, topotronics, and valleytronics. By integrating 2D vdW magnetic materials with nonmagnetic metals or insulators into heterostructures, one can realize a range of spintronic devices, including spin valves, magnetic tunnel junctions, spin field-effect transistors, and spin tunnel field-effect transistors, which serve as fundamental components for high-density information storage, advanced information processing, magnetic sensing, and non-volatile memory technologies.[18-20] Moreover, the dissipationless edge states associated with the QAHE and the topological magnetoelectric coupling in axion insulators position these materials and their heterostructures as key candidates for next-generation low-power electronic devices and magnetoelectric random access memory.[16, 21] In the context of valleytronics, magnetic exchange-induced valley splitting in 2D vdW heterostructures enables controllable valley polarization, an essential prerequisite for devices such as valley splitters,[22] valley separator[23] and controllable valley Hall effect transistor.[24, 25] Furthermore, the ability of certain 2D vdW magnets to host magnetic skyrmions, topologically protected spin textures considered promising as information carriers, opens up avenues for designing ultra-dense racetrack memory and spin-based logic devices.[26, 27]

This review aims to provide a comprehensive overview of the electronic and magnetic properties, as well as the advanced functionalities, of two-dimensional van der Waals magnetic materials and their heterostructures, with particular emphasis on the role of magnetic anisotropy and the underlying physics (Figure 1). We begin by discussing the critical role of magnetic anisotropy in stabilizing long-range magnetic order in 2D vdW systems. We then review the physical mechanisms that give rise to magnetic anisotropy across a range of 2D vdW magnetic materials, including both insulating and metallic systems, and explore how spin orientation affects their electronic, magnetic, and topological properties. Given the pivotal role of magnetic anisotropy in tuning magnetic behavior and emergent quantum phenomena, we also highlight diverse strategies for engineering anisotropy through both external and intrinsic means, including alloying, doping, strain, gating, and applied pressure. Finally, we conclude with a discussion of the current challenges and future opportunities in the study of magnetic anisotropy in 2D vdW magnets and their heterostructures. This review is specifically dedicated to magnetic anisotropy, with a focus on 2D vdW magnetic materials that have been experimentally realized, as well as their heterostructures. For broader discussions on other aspects of 2D magnetism, we refer readers to several excellent reviews.[28-34]

## 2. Importance of Magnetic Anisotropy for 2D vdW Magnets

When magnetization is rotated from its preferred orientation (i.e., the magnetic easy axis) to a direction of higher energy (i.e., the magnetic hard axis), the associated energy cost, known as the MAE, typically ranges from sub μeV to a few meV per atom. Although this energy scale is small compared to the total magnetic energy of a material, magnetic anisotropy plays a disproportionately significant role in stabilizing long-range magnetic order in 2D magnetic materials. Before delving into its role in enabling 2D magnetism, we first provide a detailed introduction to the three main components of magnetic anisotropy.

### 2.1 An Overview of Magnetic Anisotropy

For a pair of spins $S_i$ and $S_j$, the most general bilinear spin Hamiltonian can be expressed as follows:[35, 36]

$$H_{ij} = J_{ij}^{\alpha\beta} S_i^{\alpha} S_j^{\beta} \qquad (1).$$

where $\alpha$ and $\beta$ run over $x$, $y$, and $z$ axes, $J_{ij}^{\alpha\beta}$ is the exchange interaction matrix between spins at sites $i$ and $j$. This 3-by-3 matrix is generally determined by the crystal symmetries of magnetic materials and has been simplified to many different forms. For an isotropic Heisenberg spin model, $J_{ij}^{\alpha\beta}$ is simplified as $J_{ij}^{xx} = J_{ij}^{yy} = J_{ij}^{zz} = J_{ij}$ or $J_{ij}^{\alpha\beta} = J_{ij} I$, with $I$ being a 3-by-3 identity matrix. In contrast, one of the most canonical anisotropic exchange spin models is the Ising model[37] with a spin Hamiltonian of the form $H_{ij}^{\text{Ising}} = J_{ij}^{zz} S_i^z S_j^z$, which allows spin alignment only along a fixed quantization axis, typically the $z$ axis (Figure 2A). Another important example is the 2D XY model for studies of in-plane magnetic anisotropy such as the Berezinskii-Kosterlitz-Thouless (BKT) transition.[38-40] The XY model has the Hamiltonian $H_{ij}^{\text{2D-XY}} = J_{ij}\left(S_i^x S_j^x + S_i^y S_j^y\right)$ (Figure 2B). A generalized form is the 2D XXZ model, which is described by $H_{ij}^{\text{2D-XXZ}} = J_{ij}^{\perp}\left(S_i^x S_j^x + S_i^y S_j^y\right) + J_{ij}^z S_i^z S_j^z$ where the anisotropy between $J_{ij}^{\perp}$ and $J_{ij}^z$ tunes the spin interactions between easy-plane and easy-axis limit. Another unique case is the Kitaev model characterized by bond-dependent Ising-type interactions. Its spin Hamiltonian takes the form $H^{\text{Kitaev}} = -\sum_{ij\in\gamma\,\text{bonds}} K_\gamma S_i^\gamma S_j^\gamma$,[41] where the easy axis $\gamma$ depends on the spatial orientation of the $\gamma$-type bond (Figure 2C). Anisotropy can also arise from the antisymmetric DM interaction, given by $H_{ij}^{DM} = \boldsymbol{D}_{ij} \cdot \left(\boldsymbol{S}_i \times \boldsymbol{S}_j\right)$ (Figure 2D), which stems fundamentally from SOC in systems lacking inversion symmetry.[42, 43] Finally, off-diagonal symmetric exchange interactions, described by the Hamiltonian $H_{ij}^{\Gamma} = \sum_{\alpha,\beta,\gamma\in x,y,z} \Gamma_{ij}^\gamma \left(S_i^\alpha S_j^\beta + S_i^\beta S_j^\alpha\right)$, also contribute to magnetic anisotropy by introducing anisotropic spin couplings that favor specific spin orientations. These interactions have been identified in spin-orbit-entangled Kitaev materials such as $\alpha$-RuCl$_3$ and Na$_2$IrO$_3$, where they play a significant role in shaping the magnetic ground states and excitations.[44-46]

For many magnetic materials of interest, the MCA in their spin Hamiltonians can be represented by a term, $H_{\text{MCA}}$, in the following form:[2]

$$H_{\text{MCA}} = \sum_i \sum_{\alpha\beta} A_i^{\alpha\beta} S_i^\alpha S_i^\beta \qquad (2).$$

where $A_i^{\alpha\beta}$ is the MCA parameter. This form of anisotropy reflects its single-ion nature, therefore it is referred to simply as single-ion anisotropy (SIA). For 2D vdW magnetic materials with uniaxial magnetic anisotropy, the MCA is either out-of-plane or in-plane. Therefore, their MCA is often simplified by a spin Hamiltonian, $H_{\text{MCA}}^{2D} = \sum_i A_i \left(S_i^z\right)^2$, with $A_i > 0$ indicating an in-plane easy axis and $A_i < 0$ corresponding to an out-of-plane easy axis.

In all magnetic materials, dipole-dipole interactions are inherently present due to the magnetic moments associated with spins. The contribution of the magnetostatic interaction to the spin Hamiltonian is given by[2]

$$H_{dd} = \frac{1}{2}\frac{\mu_0}{4\pi}\sum_{i\neq j}\frac{1}{r_{ij}^3}\left[\boldsymbol{m}_i \cdot \boldsymbol{m}_j - \frac{3}{r_{ij}^2}\left(\boldsymbol{m}_i \cdot \boldsymbol{r}_{ij}\right)\left(\boldsymbol{m}_j \cdot \boldsymbol{r}_{ij}\right)\right] \qquad (3).$$

where the sum runs over all pair of spin sites *i* and *j*; $\boldsymbol{m}_i$ is the local magnetic moment at site *i* and $\boldsymbol{r}_{ij}$ is the vector connecting sites *i* and *j*. Clearly, the dipole-dipole interaction is long-ranged, since it decays slowly with the distance as $1/r^3$. In particular, it depends on both the relative magnetic moment orientation of two spins and their orientation with respect to $\boldsymbol{r}_{ij}$. As a result, the dipole-dipole interactions break the rotational symmetry between in-plane and out-of-plane spin orientations in a magnetically ordered 2D system, energetically favoring in-plane spin alignment over out-of-plane orientation. This effect induces a geometrically driven in-plane magnetic anisotropy, which becomes particularly significant in 2D vdW MLs where SOC is weak. Therefore, dipole-dipole interactions constitute an intrinsic source of magnetic anisotropy in 2D systems, contributing to the stabilization of specific spin textures and magnetic phases.

**2.2 Mermin-Wagner Theorem**

We now turn to the well-known Mermin–Wagner theorem, formulated in 1966, which plays a central role in understanding magnetic order in low-dimensional systems.[3] This theorem states that at finite (non-zero) temperatures, no long-range FM or AFM order can exist in one- or two-dimensional isotropic spin-*S* Heisenberg models with finite-range exchange interactions. In contrast, 3D isotropic Heisenberg models can sustain long-range magnetic order at finite temperatures, even though a complete rigorous solution is still lacking. This dimensional crossover underscores the crucial role of dimensionality in the emergence and stability of long-range magnetic order. From a physical perspective, the absence of long-range magnetic order in 2D isotropic systems arises from the proliferation of low-energy spin-wave excitations (i.e., Goldstone modes) at finite temperatures. These thermal spin fluctuations diverge in the thermodynamic limit, thereby destroying the long-range coherence of spin alignment. The Mermin–Wagner theorem therefore highlights the crucial role of symmetry-breaking perturbations, such as magnetic anisotropy, in stabilizing magnetic order in two dimensions. While this result may seem discouraging for the exploration of low-dimensional magnetic systems, it is important to emphasize that the Mermin–Wagner theorem relies on two key assumptions: (i) the exchange interactions between spins are isotropic, and (ii) the exchange interactions are short-range, decaying faster than $1/r^3$.[3, 47]

To gain a more intuitive understanding of the Mermin–Wagner theorem, let us consider an isotropic spin-*S* Heisenberg model on a 2D honeycomb spin lattice (Figure 2E). For simplicity, we set the lattice basis vectors to be unit vectors. The spin Hamiltonian of this model is given by $H = -\frac{J}{2}\sum_{\langle ij \rangle} \vec{S}_i \cdot \vec{S}_j$, where the isotropic Heisenberg exchange interaction between the nearest neighbor (NN) spin sites is FM (i.e., $J > 0$), and indices $i$ and $j$ run over the spin sites A and B. At zero temperature, the magnetic ground state of this model is a FM order with all spins aligning along the same direction (Figure 2E). At finite temperatures, thermal energy excits spins away from this aligned state. These excitations manifest as spin waves, or magnons, which are collective modes arising from the small-angle precession of spins around the ordered direction. Within the linear spin wave approximation, the energy spectrum for the spin wave (i.e. magnons) is

$$E^{\pm}(\vec{k}) = JS\left(3 \pm \sqrt{\left|e^{i\vec{k}\cdot\vec{a}_1} + e^{i\vec{k}\cdot\vec{a}_2} + e^{i\vec{k}\cdot\vec{a}_2}\right|^2}\right) \qquad (4).$$

Here, $\vec{a}_i$ ($i = 1, 2, 3$) represents the NN connection vectors and $\vec{k} = (k_x, k_y)$ is the wave vector in the reciprocal space.

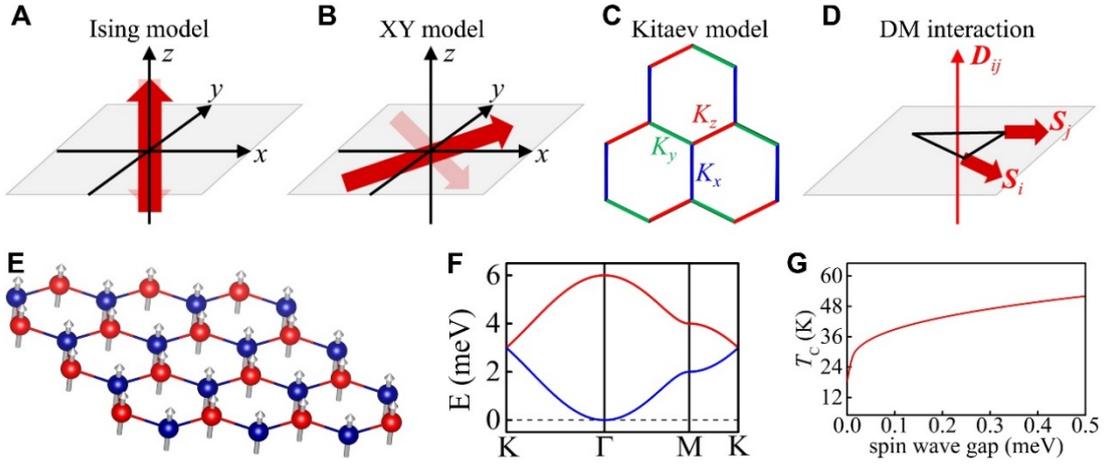

*Figure 2. Schematic illustrations of A) Ising mode, B) XY model, C) Kitaev model and D) the DM interaction. In A), B) and D), spins are depicted by the red arrows. In the Ising model, spins are preferred along the z axis. In the XY model, spins are preferred in the xy plane. In the Kitaev model, the easy axis x, y, and z bonds are indicated by the blue, green and red, respectively. E) Schematic of isotropic spin-S Heisenberg model on a 2D honeycomb spin lattice. Spin site A and B are indicated by blue and red balls, respectively. Gray arrows represent spins. F) The energy spectrum (i.e., Eq. (4)) of spin waves. Here we set $S = 1$ and $J = 1$ meV for simplicity. G) The dependence of $T_C$ on the spin wave gap based on Eq. (7). Here we set $S = 3/2$ and $J = 6$ meV.*

As shown in Figure 2F, the energy spectrum of Eq. (4) is gapless at the $\Gamma$ point. Because spin waves carry angular momenta, the correction of the saturated magnetization at temperature $T$ is

$$\Delta m(T) = \frac{g\mu_B}{2(2\pi)^2} \int_{BZ} \left(\frac{1}{e^{E^-(\vec{k})/k_BT} - 1} + \frac{1}{e^{E^+(\vec{k})/k_BT} - 1}\right) d^2\vec{k} \qquad (5).$$

In Eq. (5), $g$ and $\mu_B$ are Landé $g$-factor and Bohr magneton, respectively. Even at a very low temperature, spin waves with low energy can still be excited. Under such circumstance, the integral in Eq. (5) is controlled by the low-energy band $E^-(\vec{k})$ and this band can be replaced by its second-order expansion around the $\Gamma$ point, i.e., $E^-(\vec{k}) \cong JSk^2/4$. Besides, the integral over the first Brillouin zone can be done within a circle of radius $k_c$ with the constraint $\frac{1}{2\pi}\int_0^{k_c} k\,dk = 2$.[48] Taking these together, we obtain $\Delta m(T) = \frac{g\mu_B}{\pi}\int_0^{k_c}\frac{k_B T}{JS}\frac{dk}{k} \to \infty$. This divergence clearly indicates that the magnetization must be zero at a finite temperature. Hence, the FM order in the isotropic spin-$S$ Heisenberg model on a 2D honeycomb lattice is destroyed by thermal excitations, consistent with Mermin-Wagner theorem.[3]

By predicting the absence of long-range magnetic order in 2D isotropic spin systems at finite temperatures, the Mermin–Wagner theorem suggests fundamental limitations for the practical use of 2D magnets in spintronic applications, which typically require robust and controllable magnetic order. However, real 2D magnetic materials deviate significantly from the idealized conditions assumed by Mermin and Wagner. On one hand, the Heisenberg exchange interactions are rarely perfectly isotropic, owing to intrinsic low crystal symmetries and the presence of SOC. Notably, the SIA—an on-site contribution that favors specific spin orientations—is not accounted for in the Mermin–Wagner framework. On the other hand, dipole–dipole interactions between magnetic moments, which are long-range in nature, also fall outside the scope of the theorem. These interactions break the spin-rotational symmetry and can energetically favor certain spin alignments, such as in-plane over out-of-plane orientations. The discrepancy between the assumptions of the Mermin–Wagner theorem and the realistic physical properties of 2D magnetic materials highlights the fundamental and practical importance of exploring how long-range magnetic orders emerge in real 2D systems. Both theoretical models incorporating magnetic anisotropies and experimental discoveries of 2D magnets with robust long-range order have shown that, despite the limitations of idealized models, 2D magnetic materials can indeed host rich and technologically promising magnetic phases.

**2.3 Early Studies of 2D Ferromagnetism in Magnetic Ultrathin Films**

As early as 1960s, 2D ferromagnetism in ultrathin magnetic films has already attracted research interest. Ferromagnetism with a Curie temperature ($T_C$) of 220 K in NiFe ultrathin films with a thickness of 1.8 ML was reported by Gradmann for the first time in 1968.[4] During the period of 1980-2000, long-range FM orders with $T_C$s from several to a few hundred Kelvins was experimentally observed in Fe, Co, and Ni MLs deposited on different substrates.[30] In an angle-resolved photoemission experiment, a magnetic exchange splitting similar to that of bulk Co was observed in Co ML grown on Cu(111) in 1982,[49] indicating the FM order in Co ML. Interestingly, the saturation magnetization of the Co ML grown on Cu(111) follows the solution of the 2D Ising model in a broad range of temperatures.[50]

A key insight from experimental investigations of magnetic ultrathin films is the critical role of magnetic anisotropies in governing their magnetic phase transitions.[51] Since magnetic anisotropies primarily originate from MCA and dipole-diploe interactions, 2D Heisenberg ferromagnets incorporating one or both of them are of fundamental interest for theoretical studies. As early as 1976, Maleev demonstrated that dipole-dipole interactions which decrease like $1/r^3$ lead to the stabilization of FM orders in 2D Heisenberg ferromagnets at nonzero temperature.[52] This stabilization arises because the long-range nature of dipole-dipole interactions gives rise to a linear momentum dependence in the spin wave spectra, which eliminates the divergence of the saturated magnetization, $\Delta m(T)$. In 1986, Yafet *et al.* obtained similar results in the noninteracting spin-wave approximation.[53] Two years later, Bander and Mills proved that a phase transition to ferromagnetism always occurs in 2D Heisenberg ferromagnets with an arbitrarily small perpendicular magnetic anisotropy.[12] In 1991, Bruno discussed 2D Heisenberg ferromagnets with both dipole-dipole interactions and a uniaxial MCA using spin wave theory[47] and showed those: i) when the total magnetic anisotropy is perpendicular, the stabilization of FM orders at finite temperatures is due mainly to the perpendicular MCA induced gap at the bottom of spin wave spectra whereas dipole-dipole interactions play a negligible role; ii) when the magnetic anisotropy is in-plane, spin wave spectra are gapless so that the stabilization of the FM order is due to the long-range character of the dipole-dipole interactions. Note that the second result of Bruno's study is consistent with the previous results in the works of Maleev[52] and Yafet.[53] Clearly, both experimental and theoretical investigations into magnetic ultrathin films consistently highlight the essential role of magnetic anisotropies in enabling the long-range FM orders in 2D magnetic materials.

As a concrete example, we revisit the isotropic spin-$S$ Heisenberg model on a 2D honeycomb spin lattice to illustrate how magnetic anisotropies can stabilize long-range FM order in 2D. Within the framework of linear spin wave theory, a uniaxial magnetic anisotropy opens a finite spin wave gap, $\Delta$, which suppresses low-energy excitations and enables the emergence of long-range order at finite temperatures.[47] With this gap, the low-temperature magnetization for $\Delta/k_B T \gg 1$ is approximated by[47, 48]

$$M(T) = g\mu_B \left( S - \frac{k_B T}{2\pi JS} e^{-\Delta/k_B T} \right) \qquad (6).$$

Eq. (6) clearly shows that the presence of a spin wave gap eliminates the divergence of spin fluctuations, allowing the magnetization M(T) to persist at finite temperatures. Defining $T_C$ as the temperature at which the magnetization decreases to $M(T_C)=g\mu_B S/2$, we roughly estimate $T_C$ within a mean field approximation by the following equation:[48]

$$k_B T_C \simeq \frac{\pi JS^2}{\ln\left(\frac{\Delta + 2\pi JS}{\Delta}\right)^2} \qquad (7).$$

As shown in Figure 2G, the Curie temperature $T_C$ monotonically increases with the spin wave gap, which is determined by the MAE. Therefore, strong uniaxial magnetic anisotropy is essential for achieving high-$T_C$ ferromagnetism in 2D magnetic materials.

## 2.4 Experimental Characterization of van der Waals Magnetic Materials

Although magnetic ultrathin films have served as a valuable platform for studying 2D magnetism, their strong dependence on surface and substrate conditions often complicates the interpretation of experimental results. Recent years have seen a surge in experimental investigations into 2D vdW magnetic materials, driven by their promise for spintronic applications and fundamental studies of low-dimensional magnetism. Experimental efforts have focused on probing magnetic ordering, anisotropy, spin dynamics, and tunability under external stimuli such as strain, gating, and layer stacking. In particular, vdW materials have weak interlayer interactions and minimal environmental sensitivity, offer an ideal alternative for fundamental studies of 2D magnetism. Inspired by the discovery of graphene exfoliated from graphite using adhesive Scotch tape, [54] many magnetic vdW monolayers can similarly be obtained through mechanical exfoliation, making them both experimentally accessible and attractive for fundamental and applied research.[28-30, 32, 55] In contrast to the complexities associated with defect-induced ferromagnetism in otherwise non-magnetic 2D vdW materials such as graphene,[56-58] the magnetism in monolayer and few-layer vdW magnetic materials is intrinsic, more reproducible, and significantly more robust. These materials also exhibit higher magnetization, making them far more amenable to experimental investigation and enabling clearer insights into the fundamental physics of two-dimensional magnetism.

A major breakthrough in the study of atomically thin 2D magnetism was the successful exfoliation of the vdW transition-metal phosphorus trisulfide $FePS_3$ in 2016.[6, 7] $FePS_3$ is an AFM material of particular interest for 2D magnetism because its magnetism can be well described by a 2D Ising model on a honeycomb lattice, due to its strong out-of-plane magnetic anisotropy (Figure 3A).[59, 60] Remarkably, Ising-type

AFM order persists down to the ML limit, with a Néel temperature exceeding 100 K, as independently confirmed by two groups using Raman scattering techniques (Figure 3B-3C).[37] This discovery is fundamentally important for several reasons. First, it provides experimental verification of the long-theorized phase transition in a 2D Ising system, originally proved by Onsager in 1944.[27] Second, it offers direct evidence for intrinsic magnetic ordering in vdW magnetic materials, making a significant conceptual advancement in condensed matter physics. From the perspective of magnetic anisotropy, the 2D Ising model can be viewed as a limiting case of the 2D Heisenberg model with infinitely strong uniaxial magnetic anisotropy. Therefore, the presence of magnetic anisotropy in real 2D magnetic materials effectively violates the assumptions of the Mermin–Wagner theorem, thereby permitting the emergence of long-range magnetic order in two dimensions.

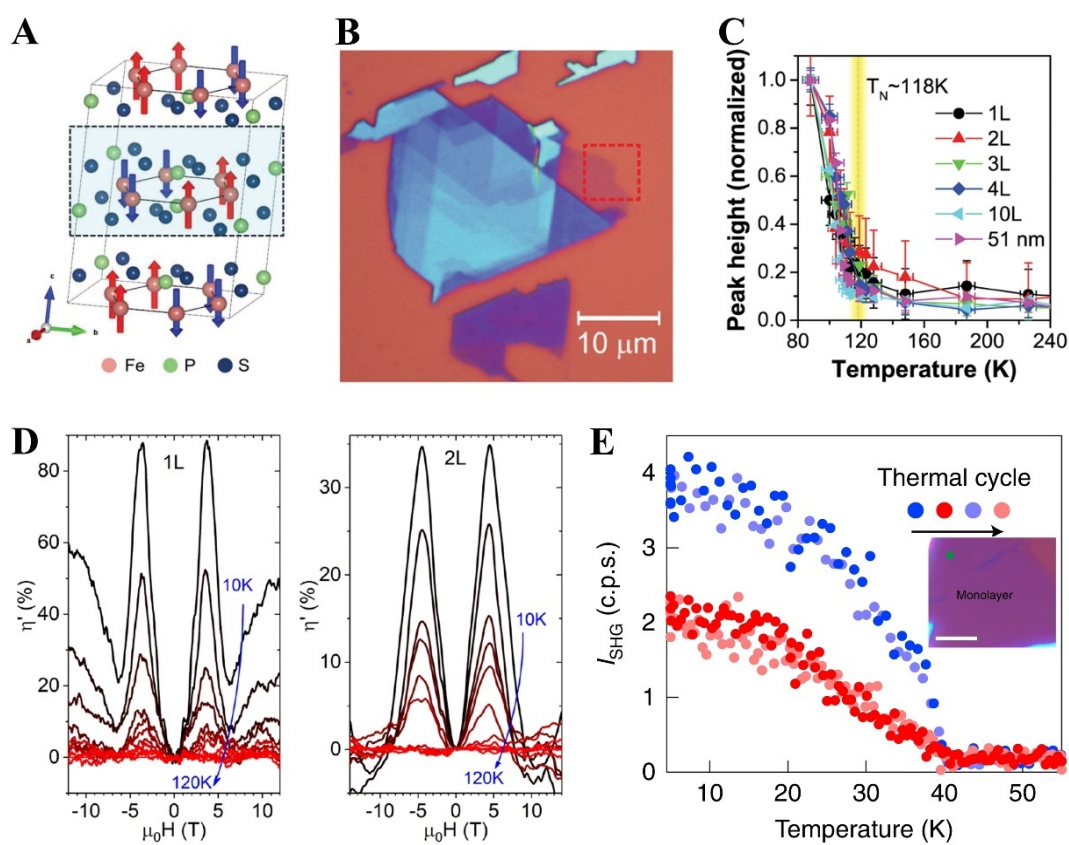

*Figure 3.* A) *The crystal structure of FePS$_3$. The sky-blue shadow highlights the ML structure of FePS$_3$. Reproduced with permission.[60] Copyright 2022, The Authors. B) Optical contrast of FePS$_3$ ML. C) The temperature dependent Raman peak height of FePS$_3$ with different thickness. B) and C) are reproduced with permission.[6] Copyright 2016, American Chemical Society. D) The tunneling magnetoresistance $\eta'(H)$ of ML (left panel) and bilayer (right panel) MnPS$_3$ as a function of magnetic field (applied out-of-plane to the layers), as temperature is increased from 10 to 120 K in 10 K steps.*

*Reproduced with permission.*[61] *Copyright 2020, American Chemical Society. E) The dependence of second-harmonic generation intensity of MnPSe$_3$ ML on temperature. Insert: the optical image of a MnPS$_3$ ML sample. Reproduced with permission.*[62] *Copyright 2021, The Author(s), under exclusive licence to Springer Nature Limited.*

Two other vdW transition-metal phosphorus trisulfides, MnPS$_3$ and NiPS$_3$, also serve as important testbeds for exploring 2D antiferromagnetism. A neutron diffraction study suggested that bulk MnPS$_3$ exhibits out-of-plane magnetic anisotropy.[63] Kim *et al.* investigated the thickness dependence of the magnetic phase transition in MnPS$_3$ using Raman spectroscopy and reported that AFM ordering remains surprisingly robust in the MnPS$_3$ bilayer.[64] This was further supported by cryogenic second-harmonic generation microscopy measurement, which confirmed a Néel-type AFM order in the MnPS$_3$ bilayer with a Néel temperature around 60 K.[65] Additionally, tunneling magnetoresistance measurements on atomically thin MnPS$_3$ crystals by Long et al. demonstrated that AFM order persists down to the monolayer limit (Figure 3D).[61] In contrast, bulk NiPS$_3$ is considered an XXZ-type antiferromagnet.[66, 67] Kim et al. examined the evolution of AFM ordering in NiPS$_3$ through layer-dependent Raman spectroscopy and concluded that while AFM order persists in the bilayer, it is significantly suppressed in the monolayer limit.[68] However, a subsequent study using helicity-resolved Raman and ultrafast spectroscopy revealed that monolayer NiPS$_3$ remains magnetically ordered, undergoing a Berezinskii–Kosterlitz–Thouless (BKT) transition at $T_{BKT} \approx 140$ K.[69] Overall, these findings demonstrate that despite dimensional reduction, two-dimensional antiferromagnetism can be sustained in van der Waals materials such as MnPS$_3$ and NiPS$_3$ due to the presence of magnetic anisotropy.

A neutron diffraction study revealed that MnPSe$_3$ exhibits strong in-plane magnetic anisotropy.[70] As a result, MnPSe$_3$ serves as an excellent platform for exploring two-dimensional magnetism described by the XY model with six-state clock order, consistent with its $S_6$ point group symmetry. Using spatially resolved second-harmonic generation, Ni et al. provided direct evidence of long-range AFM order in MnPSe$_3$ ML (Figure 3E), along with Ising-type switching of the Néel vector.[62] Remarkably, they also demonstrated that applying uniaxial strain can rotate the Néel vector to align along arbitrary in-plane directions, independent of the underlying crystal axes. This strain-induced reorientation suggests a tunable magnetic anisotropy and a change in the universality class of the magnetic phase transition, highlighting MnPSe$_3$ as a promising material for strain-engineered 2D magnetism.

For most spintronic applications[71], 2D ferromagnetism is generally more desirable[72] than antiferromagnetism. A major breakthrough in this area came in 2017 with the experimental discovery of intrinsic long-range FM order in few-layer $Cr_2Ge_2Te_6$ and monolayer $CrI_3$.[5, 73] Remarkably, Gong *et al*. demonstrated unprecedented magnetic field control over the FM transition temperature in few-layer $Cr_2Ge_2Te_6$ (Figure 4A).[73] This behavior can be understood by recognizing that $Cr_2Ge_2Te_6$ is a soft vdW magnetic material with a very small easy-axis SIA; thus an applied magnetic field effectively mimic the role of magnetic anisotropy in stabilizing long-rang FM order. A rapid increase in the FM transition temperature is observed under small magnetic fields. In contrast, $CrI_3$ is a vdW ferromagnet with strong easy-axis magnetic anisotropy.[5, 48] As a result, its robust ferromagnetism persists down to the ML limit even in the absence of magnetic field, as confirmed experimentally by Huang *et al*. (Figure 4B).[5] Furthermore, neutron scattering measurements by Chen et al. revealed that the FM phase transition in $CrI_3$ is weakly first order and governed by SOC induced magnetic anisotropy[74] rather than magnetic exchange interactions as in conventional ferromagnets. Again, these findings underscore the critical role of magnetic anisotropy in stabilizing two-dimensional ferromagnetism in atomically thin vdW magnetic materials and highlight the tunability of their magnetic properties for potential spintronic applications.

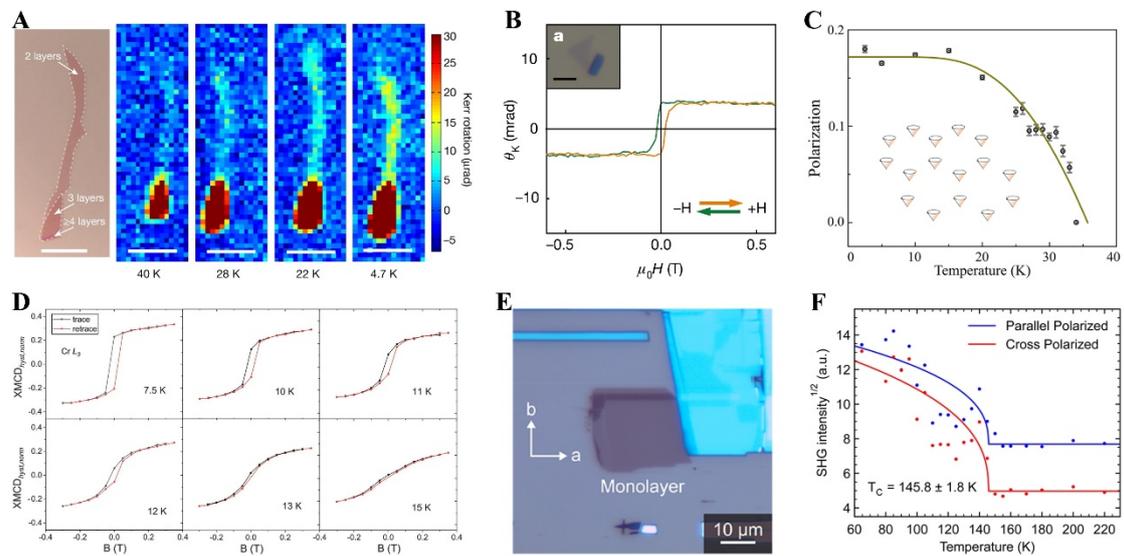

***Figure 4.*** *A) The optical image of exfoliated bilayer $Cr_2Ge_2Te_6$ and its Kerr rotation signal under 0.075 T as a function of the temperature from 4.7 to 40 K. Reproduced with permission.[73] Copyright 2017, Macmillan Publishers Limited, part of Springer Nature. B) The polar MOKE signal for an isolated $CrI_3$ ML whose optical image is shown by the inset. Reproduced with permission.[5] Copyright 2017, Macmillan Publishers Limited, part of Springer Nature. C) The polarization as a function of temperature for $CrBr_3$ ML. Reproduced with permission.[75] Copyright 2019, American*

*Chemical Society. D) The dependence of XMCD signal on an in-plane magnetic field, taken at various temperatures. Reproduced with permission.[76] Copyright 2021, AAAS. E) The optical microscopy image of a CrSBr ML on a silicon substrate. F) Square roots of the average second harmonic generation intensities of CrSBr as a function of temperature. The solid curves are fits to $(1-T/T_C)^\beta$ with $\beta$ = 0.36 to give $T_C$ = 146 ± 2 K. E) and F) are reproduced with permission.[77] Copyright 2021, American Chemical Society.*

As close relative of $CrI_3$, $CrBr_3$ is also a vdW ferromagnet.[78] Multiple experimental studies have confirmed that out-of-plane FM order can be retained in $CrBr_3$ ML.[75, 79-81] The $T_C$ of $CrBr_3$ ML is 34 K, slightly lower than the bulk value (i.e., 37 K) (Figure 4C).[75] While both $CrBr_3$ and $CrI_3$ MLs exhibit out-of-plane magnetic anisotropy, the MAE of $CrBr_3$ is about four times smaller than that of $CrI_3$.[81] This relatively weaker anisotropy makes bilayer $CrBr_3$ an ideal platform for engineering non-collinear magnetic states via twisting.[81] It is worth mentioning that the strong out-of-plane magnetic anisotropy of $CrI_3$ inhibits the formation of smooth non-collinear magnetic textures in twisted $CrI_3$ bilayer.[81]

Among the chromium trihalides $CrX_3$ ($X$ = Cl, Br, I), $CrCl_3$ ML stands out due to its in-plane magnetic anisotropy, in contrast to the out-of-plane anisotropy observed in $CrBr_3$ and $CrI_3$.[76] Bedoya-Pinto et al. demonstrated that when a single $CrCl_3$ monolayer is epitaxially grown on a graphene/6H-SiC(0001) substrate, it exhibits robust FM ordering with critical behavior consistent with a two-dimensional XY model, below a $T_C$ of 13 K (Figure 4D).[76] This behavior arises from the in-plane rotational symmetry of the magnetic anisotropy in $CrCl_3$ ML, enabling the realization of a finite-size BKT phase transition, and uncommon phenomenon in 2D vdW magnet. The emergence of XY-type magnetism in $CrCl_3$ ML opens avenues for exploring 2D superfluid-like spin transport and topological excitations.[76] In contrast, few-layer $CrCl_3$ exhibits insulating in-plane antiferromagnetic order, with a Néel temperature of approximately 17 K, highlighting a dimensionality-driven evolution in magnetic ground states.[82, 83] Comparative studies across the $CrX_3$ family reveal that magnetic anisotropy plays a pivotal role in determining the nature of their magnetic phases, dictating not only the spin alignment but also the universality class of their phase transitions.

Similarly, CrSBr ML is a FM semiconductor with in-plane magnetic anisotropy, despite its bulk being a vdW antiferromagnet with an in-plane magnetic easy axis.[84] CrSBr ML can be readily obtained via exfoliation and possess a centrosymmetric crystal structure. Remarkably, the magnetism in CrSBr ML can be detected through magnetic-dipole-induced second harmonic generation (SHG), which enables probing of symmetry-breaking magnetic order in centrosymmetric systems. Using SHG, Lee *et al*. demonstrated the onset of FM ordering in CrSBr ML with a $T_C$ of 146 ± 2 K (Figure 4E-4F).[77] By investigating the square roots of the average SHG intensities as a function of temperature, they found that the FM phase is best described by the anisotropic Heisenberg model, rather than the Ising or XY models (Figure 4F). This anisotropic ferromagnetic order is likely driven by a combination of single-ion anisotropy (SIA) and anisotropic exchange interactions.[84] These findings highlight CrSBr ML as a rare example of a high-$T_C$ 2D magnetic semiconductor with complex anisotropic spin interactions, opening new opportunities for exploring magneto-optical phenomena and spintronic applications in centrosymmetric layered materials.

Vanadium triiodide ($VI_3$) is a vdW FM Mott insulator with a $T_C$ of approximately 50 K in its bulk [85, 86] and is identified as a hard ferromagnet with a significant coercivity.[87] In contrast to the Ising ferromagnetism of $CrI_3$, the magnetic easy axis in $VI_3$ is not perpendicular to the layers; instead, it is canted by about 40° from the normal to the *ab* plane.[88] This complex magnetic anisotropy is likely rooted in the material's low-symmetry crystal structure and the presence of bond-dependent Kitaev interactions.[88, 89] Like $CrI_3$, $VI_3$ retain robust ferromagnetism down to the ML limit.[90] However, a surprising deviation from typical vdW magnets has been observed: Lin et al. reported an anomalous increase in $T_C$ with decreasing layer number, reaching a maximum of 60 K in the $VI_3$ ML, a reversal of the more commonly observed trend of suppressed magnetism in ultrathin limits.[90] This unusual thickness dependence may be linked to differences in stacking order between monolayer and bulk, particularly the loss of inversion symmetry, as well as the material's intricate magnetic anisotropy. [90]

Among 2D insulating vdW magnetic materials, $MnBi_2Te_4$ stands out as a unique system that intrinsically combines long-range magnetic order with nontrivial topological electronic properties.[91-93] In its bulk form, $MnBi_2Te_4$ is an AFM topological insulator (TI), exhibiting A-type AFM order with out-of-plane magnetic anisotropy and a Néel temperature of 24 K.[94, 95] Upon thinning to the 2D limit, $MnBi_2Te_4$ thin films display a unique thickness-dependent topological properties.[96] Remarkably, a zero-field QAHE has been observed in five-septuple-layer $MnBi_2Te_4$ specimen at 1.4 K, signaling the realization of a topologically nontrivial magnetic Chern insulator state in a stoichiometric compound without external doping or magnetic proximity.[97] In contrast, six-septuple-layer $MnBi_2Te_4$ thin films have been shown to host a robust axion insulator state, characterized by a gapped surface spectrum in the presence of time-reversal-breaking AFM order, but without net chiral edge conduction.[98] In both cases, the presence of strong out-of-plane magnetic anisotropy is critical, as it ensures the stability of the spin alignment necessary for breaking time-reversal symmetry—an essential ingredient for realizing these topological phases in thin films.

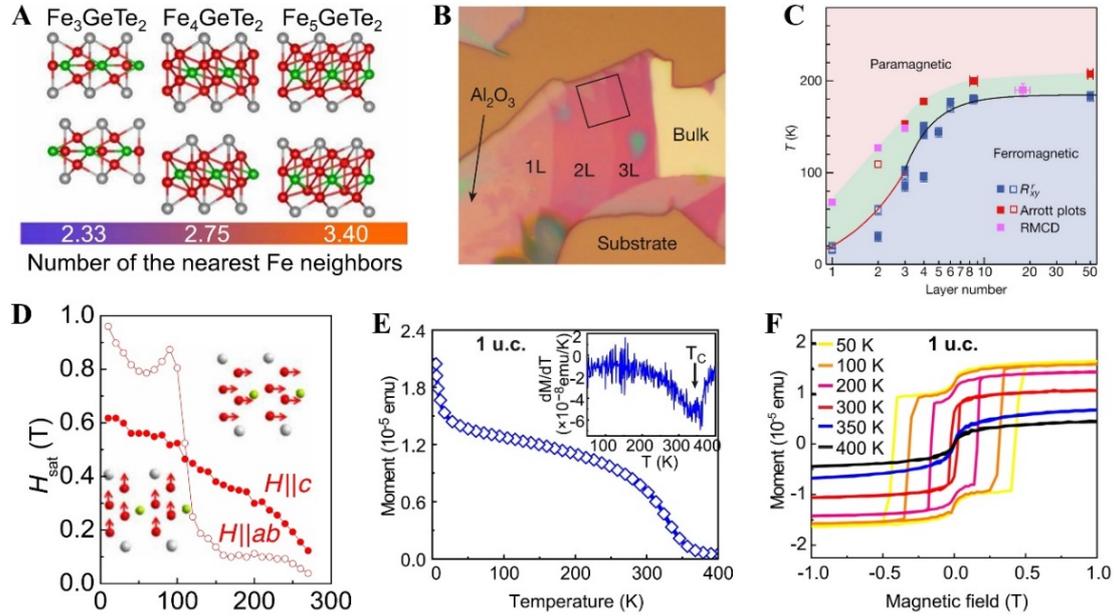

***Figure 5.*** *A) Side view of crystal structures of $Fe_nGeTe_2$ (n = 3, 2, 3). B) Optical image of typical few-layer $Fe_3GeTe_2$ flakes on the top of an $Al_2O_3$ thin film. C) Phase diagram of $Fe_3GeTe_2$ as its layer number and temperature vary. B) and C) are reproduced with permission.[99] Copyright 2018, Springer Nature Limited. D) Temperature dependence of the magnetic saturation field for H||ab (open symbols) and H||c (solid symbols) in $Fe_4GeTe_2$ single crystal. The magnetic easy axis changes from the easy axis (||c) to the easy plane (||ab) around $T_{SRT}$ = 110 K. A) and D) are reproduced with permission.[100] Copyright 2020, The Authors, some rights reserved; exclusive licensee AAAS. E) The temperature-dependent magnetization of $Fe_3GaTe_2$ trilayer under an out-of-plane magnetic field. F) M-H curves of $Fe_3GaTe_2$ trilayer at varying temperature under out-of-plane magnetic field. E) and F) are reproduced with permission.[101] Copyright 2024, the author(s).*

Nevertheless, the $T_C$ of most 2D insulating magnetic materials remain relatively low, limiting their practical utility in spintronic devices that require robust magnetism at higher temperatures. Under such circumstances, FM vdW metals are naturally noticed as they often exhibit significantly higher $T_C$. A prototypical example is $Fe_3GeTe_2$, a metallic vdW ferromagnet with a bulk $T_C$ of about 220 K,[102] making it a promising candidate for realizing high-temperature 2D ferromagnetism. From a theoretical standpoint, Zhuang *et al*. employed density-functional theory to predict that $Fe_3GeTe_6$ ML is a stable 2D Stoner ferromagnet with strong uniaxial MCA which favors robust out-of-plane spin alignment.[103] Experimentally, Fei *et al.* demonstrated that mechanically exfoliated Fe$_3$GeTe$_2$ ML exhibit long-range FM order with a $T_C$ of 130 K,[104] while Deng *et al.*, using an Al$_2$O$_3$-assisted exfoliation technique, reported a lower $T_C$ of around 68 K in monolayer samples (Figure 5B-5C).[99] Despite the discrepancy in Curie temperatures reported by these independent studies—likely due to variations in sample quality, strain, or substrate effects—both groups confirmed the presence of strong out-of-plane magnetic anisotropy in Fe$_3$GeTe$_2$ ML.[99, 104] Inspired by these findings, Fe$_3$GeTe$_2$ has emerged as a model system for investigating tunable two-dimensional metallic ferromagnetism, offering promising opportunities for the development of spintronic devices capable of operating at elevated temperatures.

As changes in layer number and stacking significantly impact both crystal structure and magnetic properties of vdW magnets, it is instructive to compare the magnetic behaviors across the family Fe$_n$GeTe$_2$ ($n$ = 3, 4, and 5). Figure 5A depicts the crystal structures of Fe$_n$GeTe$_2$, highlighting the progressive increase in Fe layers leading to a corresponding rise in the number of nearest-neighbor Fe atoms. This structural evolution is closely tied to the enhancement of magnetic ordering temperatures: the measured Curie Temperatures are about 220 K, 270 K and 310 K for Fe$_3$GeTe$_2$, Fe$_4$GeTe$_2$, and Fe$_5$GeTe$_2$ thin flakes, respectively.[102, 105, 106] However, Fe$_n$GeTe$_2$ exhibits a different magnetic anisotropy when $n$ goes from 3 to 5. First, Fe$_3$GeTe$_2$ is a hard magnet with a strong out-of-plane magnetic anisotropy.[107] In contrast, Fe$_4$GeTe$_2$ has a temperature-dependent magnetic anisotropy. As shown in Figure 5D, its magnetic easy axis lies in the *ab*-plane at high temperature but undergoes a spin reorientation transition (SRT) to the out-of-plane *c* axis below a critical temperature ($T_{SRT}$) of 110 K.[14, 100, 105, 108, 109] Notably, in bilayer Fe$_4$GeTe$_2$, nitrogen-vacancy magnetometer measurements reveal a lower SRT temperature of approximately 80 K,[110] indicating dimensionality and thickness effects on the spin reorientation dynamics. Finally, Fe$_5$GeTe$_2$ introduces further complexity, exhibiting thickness-dependent magnetic behavior. In its bulk form, Fe$_5$GeTe$_2$ is a soft ferromagnet with either in-plane[111, 112] or out-of-plane[113-115] magnetic anisotropy, varying between samples. Upon exfoliation, thin Fe$_5$GeTe$_2$ flakes transition to a hard FM state, while the monolayer displays spin-glass-like behavior, a hallmark of frustrated magnetic interactions and disorder.[111] Interestingly, bilayer Fe$_5$GeTe$_2$ grown via molecular beam epitaxy (MBE) retains a FM phase with a high $T_C$ of 229 K,[116] underscoring the tunability of magnetic properties via growth technique and dimensional control. The contrasting behaviors, from robust 2D ferromagnetism in Fe$_3$GeTe$_2$ ML to spin-glass-like magnetism in Fe$_5$GeTe$_2$ ML, highlight how anisotropy and structural dimensionality jointly dictate the emergence and nature of magnetism in vdW magnets.

Fe$_3$GaTe$_2$ is another promising vdW ferromagnet incorporating iron, notable for its above-room-temperature $T_C$ up to the range of 350-380 K)[34,117]. Compared to Fe$_3$GeTe$_2$, Fe$_3$GaTe$_2$ possesses a stronger perpendicular magnetic anisotropy.[118] A perpendicularly large magnetic anisotropy constant $K_U$=6.7×10$^5$ J/m$^3$ has been reported at 300 K in a nine-unit-cell Fe$_3$GaTe$_2$ thin film.[101] Remarkably, even in the ultra-thin limit, the material retains robust magnetic order. In a one-unit-cell (i.e., trilayer) Fe$_3$GaTe$_2$ film, the perpendicular magnetic anisotropy remains strong ($K_U$=1.8×10$^5$ J/m$^3$ at 300 K), supporting FM ordering with a $T_C$ up to 345 K (Figure 5E-5F).[101] Furthermore, Wang *et al*. demonstrated that Fe$_3$GaTe$_2$ ML exhibits 2D hard ferromagnetism, with a record-high $T_C$ of 240 K among known intrinsic FM vdW MLs, based on Hall resistance and magnetoresistance measurements.[119] Fundamentally, the stabilization of 2D FM order in both Fe$_3$GeTe$_2$ and Fe$_3$GaTe$_2$ MLs can be attributed to the interplay of strong magnetic anisotropy and enhanced exchange interactions.

**Table 1.** A summary of the magnetic properties and anisotropies of 2D vdW magnetic materials which are reviewed here. MGS, $T_M$ and SL stand for magnetic ground state, magnetic transition temperature and septuple layer, respectively.

| Materials | Thickness | MGS | Magnetic anisotropy | $T_M$ (K) | Reference |
|---|---|---|---|---|---|
| FePS$_3$ | ML | AFM | Ising-like | ~ 100 | [6, 7] |
| MnPS$_3$ | ML | AFM | out-of-plane | 78 ± 5 | [61] |
| MnPSe$_3$ | ML | AFM | in-plane | 40 | [62] |
| Cr$_2$Ge$_2$Te$_6$ | bilayer | FM | weak out-of-plane | ~ 30 | [73] |
| CrI$_3$ | ML | FM | strong out-of-plane | 45 | [5] |
| CrBr$_3$ | ML | FM | out-of-plane | 34 | [75] |
| CrCl$_3$ | ML | FM | In-plane | 13 | [76] |
| CrSBr | ML | FM | In-plane | 146 | [77] |
| VI$_3$ | ML | FM | strong out-of-plane | 60 | [90] |
| MnBi$_2$Te$_4$ | six-SL | AFM | out-of-plane | 20 | [98] |
| Fe$_3$GeTe$_2$ | ML | FM | strong out-of-plane | 68,130 | [99] |
| Fe$_4$GeTe$_2$ | thin flake | FM | temperature-dependent | 270 | [105] |
| Fe$_5$GeTe$_2$ | bilayer | FM | out-of-plane | 229 | [111, 116] |
| Fe$_5$GeTe$_2$ | ML | spin-glass-like | -- | -- | [111] |
| Fe$_3$GaTe$_2$ | trilayer | FM | strong out-of-plane | 345 | [101] |
| Fe$_3$GaTe$_2$ | Few-layer | FM | strong out-of-plane | 350~380 | [117] |
| Fe$_3$GaTe$_2$ | ML | FM | strong out-of-plane | 240 | [119] |
| 1T-CrTe$_2$ | ML | FM | out-of-plane | 200 | [120] |
| α-RuCl$_3$ | ML | AFM | out-of-plane | ~ 8 | [121] |

1T-CrTe$_2$ ML is an intriguing 2D vdW magnetic materials whose magnetic ground state is highly sensitive to its in-plane lattice constant. Sun *et al*. demonstrated that FM order can persist above 300 K in ultra-thin 1T-CrTe$_2$ films.[122] Using MBE, Zhang *et al*. grew 1T-CrTe$_2$ ML on bilayer graphene and revealed a FM ground state with out-of-plane magnetic anisotropy and a $T_C$ of about 200 K.[120] Similarly, epitaxial growth of a one unit-cell thick 1T-CrTe$_2$ film on ZrTe$_2$ substrates results in a 2D ferromagnet exhibiting a clear anomalous Hall effect, indicative of robust FM ordering.[123] However, contrasting behavior is observed when 1T-CrTe$_2$ ML is grown on a SiC-supported bilayer graphene, where a stable AFM order with moderate magnetic anisotropy emerges.[124] These substrate-dependent magnetic states are consistent with theoretical predictions showing that the magnetic ground state of 1T-CrTe$_2$ ML is strongly dependent on its in-plane lattice constant. [125] Specifically, when the in-plane lattice is constrained to match that of ZrTe$_2$, the system favors a FM state with out-of-plane magnetic anisotropy, aligning well with the experimental observations.[123,126]

*α*-RuCl$_3$, a vdW magnetic material based on a 4*d* transition metal, offers a distinct contrast to the more commonly studied 3d transition metal-based vdW magnets, making it a compelling platform for exploring 2D magnetism in the context of strong spin-orbit coupling and electronic correlations. Notably, *α*-RuCl$_3$ is widely considered as a promising candidate for realizing the long-sought Kitaev quantum spin liquid, owing to its honeycomb lattice and anisotropic exchange interactions.[45, 46, 127] With the rapid progress in 2D vdW magnetic materials research, *α*-RuCl$_3$ ML has attracted increasing interest. DFT calculations predicted that introducing free carriers and optically excited electron-hole pairs can drive *α*-RuCl$_3$ ML from a proximate spin-liquid phase into a stable FM state.[128] Biswas *et al*. showed via DFT calculations that interfacing *α*-RuCl$_3$ ML with graphene can induce an insulator-to-metal transition, open the door to realizing metallic and even exotic superconducting states in its engineered heterostructures.[129] Furthermore, it is reported in an experiment that the large work function and narrow bands of *α*-RuCl$_3$ monolayer and bilayer enable the modulation doping of the exfoliated single-layer and bilayer graphene, highlighting its utility in vdW heterostructure engineering.[130] A particular interesting discovery is the reversal of magnetic anisotropy in *α*-RuCl$_3$ ML, driven by structural symmetry-breaking.[121] While the spin direction in bulk *α*-RuCl$_3$ is 35º away from the *ab* plane,[131] Yang *et al*. found that picoscale structural distortions in the *α*-RuCl$_3$ monolayer alter the off-diagonal symmetric exchange interactions, leading to a reversal from in-plane to easy-axis magnetic anisotropy.[121] The profound influence of dimensionality and subtle lattice distortions on magnetic behavior of α-RuCl$_3$ films suggests that they may serve as model systems for probing the intricate interplay between lattice symmetry, spin orbit coupling, and electron correlations in the quest for novel quantum states in vdW magnetic materials based on heavy transition metals.

Based on experimental studies of various 2D vdW magnetic materials (see the summary in Table 1), it is evident that magnetic anisotropy is not only essential for stabilizing long-range magnetic order at finite temperatures, but also plays a critical role in determining magnetic transition temperatures and enabling diverse functionalities. The rapid and ongoing expansion of this field continues to yield an increasing array of novel vdW magnetic materials and engineered heterostructures, which offer unprecedented opportunities to probe the interplay between magnetism, topology, and reduced dimensionality, opening pathways toward the realization of next-generation spintronic, topological, and quantum devices. In this context, it is increasingly important to uncover the physical origins of magnetic anisotropy across different material platforms and to develop effective strategies for its precise control and manipulation, which are essential for advancing future applications in spintronics and quantum information technologies.

## 3. Mechanisms of Magnetic Anisotropy in 2D vdW Magnets

In this section, we review the underlying mechanisms responsible for magnetic anisotropy in 2D vdW magnets. While it is widely recognized that spin–orbit coupling plays a central role in determining magnetic anisotropy, a detailed understanding of how SOC influences anisotropic magnetic behavior in 2D vdW systems remains essential. Clarifying this connection not only deepens our insight into the origin of magnetic anisotropies at the atomic scale but also informs the design and control of emergent magnetic phenomena in low-dimensional materials.

### 3.1 Magnetic Anisotropy by Ligand SOC

As one of the prototypical 2D vdW magnetic materials, $CrI_3$ ML has attracted widespread interest following the experimental discovery of long-range 2D FM order, particularly with regard to strategies for enhancing its Curie temperature.[5] Through DFT calculations and spin model Hamiltonian simulations, Lado *et al*. investigated the origin of the magnetic anisotropy of $CrI_3$ ML.[48] They identified the dominant source of magnetic anisotropy as anisotropic symmetric superexchange interactions, primarily mediated by the SOC of the iodine ligands. Their analysis further revealed that the single-ion anisotropy of the Cr local moments is minimal, suggesting that the magnetic behavior of $CrI_3$ ML is better captured by an XXZ-type Hamiltonian rather than an Ising model.[48] Although their model simulations underestimated $T_C$ by 20% compared to the experimental value based on their results, this study highlighted the crucial role of ligand SOC in generating strong out-of-plane magnetic anisotropy in $CrI_3$ ML (Figure 6A). Kim *et al*. also examined the influence of *p*-orbital SOC on the magnetic anisotropy in $CrI_3$, and independently arrived at an anisotropic XXZ spin Hamiltonian that accounts for the observed magnetic anisotropy.[132] The same conclusion has also been obtained by other first-principles studies. Yang *et al*. assigned the MAE mainly to contributions from iodine atoms, up to 0.71 meV/atom, [133] comparable to the MAE of metallic Fe atoms at the Fe/MgO interface.[134] Through detailed analysis of the density of states and orbital-resolved MAE within second-order perturbation theory, they identified that the dominant contribution arises from the difference in matrix elements between same spin $p_y$ and $p_x$ orbitals of iodine atoms (Figure 6B). In an independent first-principles investigation of the tunability of magnetic anisotropy in CrI$_3$ ML, Kim et al. similarly concluded that the strong SOC of iodine atoms is the principal source of the observed MAE. [135]

Given the strong SOC of iodine atoms, various forms of anisotropic exchange interactions have been proposed to account for the magnetic anisotropy observed in CrI$_3$ ML. By calculating exchange coupling matrix and SIA parameters using first-principles calculations, Xu *et al*. argued that the magnetic anisotropy of CrI$_3$ ML is determined by the interplay between SIA and Kiteav-like interactions.[136] They demonstrated that both SIA and Kiteav-like interactions are induced by the strong SOC of the heavy iodine atoms. Expanding on this perspective, Stavropoulos et al. considered the effects of the slight trigonal distortion in the CrI$_6$ octahedra, in conjunction with the strong ligand SOC, and derived a comprehensive spin model that includes bond-dependent Kitaev interactions, off-diagonal symmetric exchange terms ($\Gamma$ and $\Gamma'$), and SIA, in addition to the conventional isotropic Heisenberg exchange. [137] Their results revealed that the FM $\Gamma$, $\Gamma'$, and SIA terms all contribute cooperatively to the observed out-of-plane magnetic anisotropy in CrI$_3$ MLs.

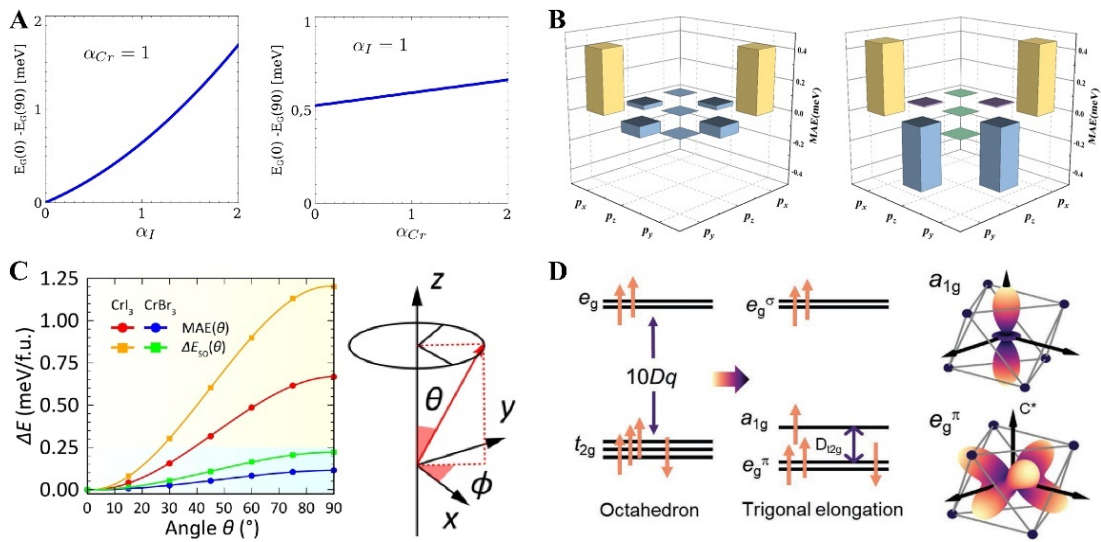

*Figure 6. A) Left panel: the dependence of MAE on the SOC of iodine atoms with the real SOC of Cr atoms. Right panel: same as the left panel but reverting the roles of*

*iodine and Cr atoms. Reproduced with permission.[48] Copyright 2017, IOP Publishing. B) Orbital-resolved MAEs of different iodine atoms in CrI$_3$ ML. Reproduced with permission.[133] Copyright 2018, American Chemical Society. C) Left panel: MAE and SOC energy as a function of the spin alignment angle with respect to the out-of-plane direction in CrI$_3$ and CrBr$_3$ MLs. Right panel: schematic of the spin alignment angle. Reproduced with permission.[138] Copyright 2021, American Physical Society. D) The high spin configuration (left panel) and $a_{1g}$ and $e_g^\pi$ states (right panel) of $Fe^{2+}$ ion in FePS$_3$ under the crystal field. Reproduced with permission.[60] Copyright 2022, The authors.*

The importance of the strong ligand SOC from iodine atoms in governing the magnetic anisotropy of CrI$_3$ is also substantiated, to some extent, by experimental evidence. By comparing the experimental Cr $L_{2,3}$ X-ray magnetic circular dichroism (XMCD) spectra with theoretical simulations based on a hybridized ground state, Frisk *et al.* concluded that Cr-I bonds in bulk CrI$_3$ have a strongly covalent character, suggesting significant hybridization between Cr *d*- and I *p*-orbitals.[139] Furthermore, Lee et al., through a combination of angle-dependent ferromagnetic resonance (FMR) measurements on high-quality CrI$_3$ single crystals and a symmetry-based theoretical analysis, identified the bond-dependent Kitaev interaction as the dominant anisotropic exchange term, while also noting that the off-diagonal symmetric exchange (Γ) is relatively small but essential for opening the observed spin-wave gap.[140] Complementary magnetic spectroscopy experiments further revealed a sizable orbital magnetic moment associated with the iodine atoms, providing direct experimental evidence of their contribution to the magnetic anisotropy in CrI$_3$.[141] Collectively, these findings support the growing consensus that the strong SOC of iodine ligands plays a central role in determining the nature and strength of magnetic anisotropy in CrI$_3$.

The attribution of magnetic anisotropy to contributions from the halogen ligands is further substantiated by comparative studies across the CrX$_3$ series (X = Cl, Br, I), which systematically highlight the role of ligand SOC strength. One interesting example is CrCl$_3$ ML. Early first-principles studies[142, 143] predicted an out-of-plane magnetic easy axis for CrCl$_3$ ML, contradicting to the experimentally observed in-plane ferromagnetism.[76] In a comprehensive DFT study, Xue *et al.*[144] calculated the SOC-induced out-of-plane MCA energies for CrCl$_3$, CrBr$_3$ and Cr$_3$I MLs to be 18, 157 and 655 μeV/Cr, respectively, reflecting the increasing SOC strength from Cl to I. Meanwhile, they evaluated the in-plane magnetic shape anisotropy arising from dipole–dipole interactions, finding comparatively small values of 58, 47, and 37 μeV per Cr atom for CrCl$_3$, CrBr$_3$ and CrI$_3$ MLs, respectively. Despite the overall weakness of dipolar interactions, their relative importance becomes apparent in CrCl$_3$, where the weak ligand SOC leads to an MCA smaller than the MSA, resulting in net in-plane magnetic anisotropy as observed in experiments.[76] In contrast, CrI$_3$ and CrBr$_3$ MLs retain out-of-plane magnetic anisotropy due to the strong SOC of Br and I atoms. This study underscores that while ligand SOC primarily governs magnetic anisotropy in heavier halides, dipolar interactions may become decisive in 2D vdW magnets composed of lighter elements. Notably, Gudelli et al. independently reported consistent findings regarding the origin of magnetic anisotropy in CrI$_3$ ML.[145]

Although bromine has a weaker SOC than iodine, its SOC strength still exceeds that of chromium. As a result, the magnetic anisotropy in CrBr$_3$ monolayers exhibits qualitative similarity to that of CrI$_3$ monolayers.[133] However, several notable differences distinguish their magnetic anisotropy characteristics.[138] First, the MAE of CrI$_3$ ML is about five times larger than that of CrBr$_3$ ML (Figure 6C).[138] Second, CrBr$_3$ ML has a much smaller out-of-plane exchange anisotropy compared to CrI$_3$ ML.[138] Third, the two materials respond differently to biaxial strain: the MAE of CrBr$_3$ increases under tensile strain and decreases under compressive strain, following a nearly linear trend, while CrI$_3$ MLs shows an anomalous response, with MAE increasing under both compressive and tensile strain.[138] These contrasting behaviors highlight the subtle dependence of magnetic anisotropy on the SOC strength of ligand $p$ orbitals and demonstrate how small changes in chemical composition can lead to significant variations in the magnetic properties of 2D vdW magnets.[138]

The mechanism underlying magnetic anisotropy in Cr$_2$Ge$_2$Te$_2$ appears to be less understood compared to that in CrX$_3$ (X = I, Br, Cl) MLs. Experimental studies have consistently shown that Cr$_2$Ge$_2$Te$_2$ exhibits weak out-of-plane magnetic anisotropy in both bulk and few-layer forms.[73, 146, 147] From a mean-field perspective, the out-of-plane SIA in bulk Cr$_2$Ge$_2$Te$_2$ has been estimated to be approximately 0.02 meV per Cr atom, a value comparable to 0.05 meV/Cr obtained via DFT calculations with a Hubbard $U$ = 0.5 eV.[73] However, using the same $U$ parameter and a more comprehensive spin Hamiltonian, Xu et al. found that the SIA in Cr$_2$Ge$_2$Te$_2$ ML favors in-plane magnetization.[136] They argued that the weak magnetic anisotropy of Cr$_2$Ge$_2$Te$_2$ ML arise from a near-cancellation between SIA and anisotropic Kitaev-type A theoretical study by Wang et al., based on a minimal tight-binding model constructed from maximally localized Wannier functions, also supports the in-plane magnetic anisotropy in Cr$_2$Ge$_2$Te$_2$.[148] They argued that the behavior of MAE primarily originates from the Te 5p orbitals, which play a crucial role in enhancing both the ferromagnetic exchange and the Dzyaloshinskii–Moriya interactions between Cr spins. By incorporating both MCA and MSA, Yang et al. performed DFT calculations showing that out-of-plane magnetization is favored in bilayer, trilayer, and bulk Cr$_2$Ge$_2$Te$_6$, while monolayers favor in-plane magnetization. [149] This behavior arises because the relatively large in-plane MSA overweighs its MCA, a trend reminiscent of that observed in CrCl$_3$. [144] Another DFT study also found in-plane anisotropy for the Cr$_2$Ge$_2$Te$_6$ monolayer.[150] Given the well-documented shortcomings of the DFT+U approach in accurately capturing the bandgap of bulk Cr$_2$Ge$_2$Te$_6$, Menichetti *et al.* employed hybrid functional calculations and achieved good agreement with experimental bandgaps measured via angle-resolved photoemission spectroscopy (ARPES). Their results underscore the critical role of nonlocal electron–electron interactions in governing the electronic properties of Cr$_2$Ge$_2$Te$_6$.[151] As a result, they found an out-of-plane SIA in Cr$_2$Ge$_2$Te$_6$ ML; however, its magnitude was smaller than the in-plane MSA, leading to a preference for in-plane magnetization. These studies collectively highlight that the competition between out-of-plane SIA and in-plane MSA is a common and critical factor in determining the magnetic easy axis in 2D vdW ferromagnets.

Since Te is positioned near iodine in the same row of the periodic table, it also has a very strong SOC. This prompts a natural question of whether the SOC of Te plays a comparable role in determining the magnetic anisotropy of Cr$_2$Ge$_2$Te$_6$ like the case of CrI$_3$. Indeed, Wang *et al.*[148] demonstrated that the Te 5$p$ states are largely responsible for the magnetic anisotropy in Cr$_2$Ge$_2$Te$_6$ ML. Using atomic- and orbital-resolved magnetic anisotropy analysis based on second-order perturbation theory, Liu *et al.* found that the MCA of Cr$_2$Ge$_2$Te$_6$ ML primarily originates from the hybridized Te-$p_y$ and Te-$p_z$ orbitals.[150] Interestingly, a first-principles study combined with machine learning unveiled that, across many 2D A$_2$B$_2$X$_6$ materials, the strength of magnetic anisotropy is dominated by the nonmagnetic X-site chalcogen, while the magnetic A-site transition metal contributes relatively little.[152] This emphasizes the crucial role of Cr-Te orbital hybridization in transferring the strong SOC of Te to the localized Cr d-orbitals, thereby playing a key role in determining the magnetic anisotropy in Cr$_2$Ge$_2$Te$_6$. Giving the rapid advancements in applying machine learning to DFT studies in materials sciences,[153] leveraging these techniques offers a promising avenue to unravel the mechanisms underlying magnetic anisotropies in 2D vdW magnetic materials.

Like Cr$_2$Ge$_2$Te$_6$, the out-of-plane magnetic anisotropy in 1T-CrTe$_2$ monolayers is primarily attributed to the strong spin–orbit coupling (SOC) of Te atoms. It is suggested that the out-of-plane magnetic anisotropy in 1T-CrTe$_2$ ML is primarily attributed to the strong SOC of Te atoms. Through systematic DFT investigations of the structural, electronic, and magnetic properties of various CrTe$_2$ phases, Liu *et al.* demonstrated that the magnetic anisotropy of 1T-CrTe$_2$ ML stems from the SOC of the Te atoms combined with superexchange coupling between the Cr-3$d$ and Te-5$p$ orbitals.[154] An atom- and orbital-resolved analysis of the MAE, based on second order perturbation theory, further confirmed that the SOC of Te atoms is the dominant contributor to the MAE in 1T-CrTe2 ML.[155] Interestingly, the magnetic anisotropy arises from a delicate competition between positive contributions due to $p_y/p_z$ orbital hybridization and negative contribution from $p_x/p_y$ hybridizations, enabling strain-induced tuning of the easy axis from out-of-plane to in-plane in 1T-CrTe2 ML.[155] Supporting this conclusion, two additional DFT studies also attributed the out-of-plane magnetic anisotropy in 1T-CrTe2 ML chiefly to the strong SOC of Te atoms, which significantly exceeds that of Cr.[156, 157]

Ultrathin MnBi2Te4 and MnBi2Se4 films are topological vdW magnetic materials whose magnetic anisotropies are generally weak. This behavior is largely attributed to the nature of the $Mn^{2+}$ magnetic ion, which has a half-filled 3d shell. DFT calculations by Li *et al.* found that MnBi2Te4 ML has nearly isotopic exchange interactions, as a result of the weak *p-d* hybridization between Mn and Te atoms.[158] The SIA of MnBi2Te4 ML cannot be explained solely by the SOC of Mn atoms. Instead, it involves the SOC of the ligand Te atoms, which alters the local Mn 3d electronic states and contributes to the observed anisotropy. It is worth noting that although Bi atoms possess strong SOC, their contribution to the magnetic anisotropy is minimal.[158] Instead, Bi atoms play a crucial role in mediating the FM exchange interactions in MnBi2Te4 ML.[159]

Although MnPS3 and MnPSe3 are isostructural and have similar magnetic moments, critical temperatures and share the same AFM ordering, they exhibit markedly different magnetic anisotropies.[160, 161] Early experimental measurements indicated that MnPS3 has a weak out-of-plane magnetic anisotropy arising from a competition between SIA and MSA.[162, 163] More recent investigations into the system's critical behavior across a wide temperature range revealed a crossover from an isotropic Heisenberg AFM state at high temperatures to a 2D XY phase at temperatures.[164-167] This intriguing spin-related critical transition in MnPS3 remains an interesting open question in the field. Theoretically, DFT calculations consistently show that MnPS3 is well characterized by isotropic Heisenberg exchange interactions and possesses only very small SIA.[168-170] In contrast, MnPSe3 exhibits unusually large XY anisotropy.[161] Given that $Mn^{2+}$ ions in their high spin $^6S$ ground state typically show minimal zero-field splitting, the zero-field splitting for the orbital singlet of $Mn^{2+}$ ions is usually small, the substantial in-plane anisotropy in MnPSe3 has been attributed to the stronger ligand SOC of the heavier Se atoms compared to S in MnPS3.[161] Analysis of zero-wavevector magnon excitations by Jana *et al.* indicated that the spins in MnPSe3 align along specific in-plane crystal axes, though these axes remain experimentally undetermined and merit further study.[171] Supporting these observations, DFT calculations confirmed that SOC induces in-plane magnetic anisotropy in MnPSe$_3$ monolayer.[172] Furthermore, chalcogen substitution in MnPS$_{3-x}$Se$_x$ (0≤$x$≤3) has been shown to effectively tune the magnetic anisotropy between out-of-plane and in-plane orientations. This tunability is primarily attributed to the enhanced SIA arising from the increasing ligand SOC contribution as Se content increases.[169]

CrSBr features an orthorhombic crystal structure and exhibits a triaxial magnetic anisotropy; explicitly, its easy, intermediate, and hard magnetic axes are along the crystallographic *b*, *a*, and *c* axes, respectively.[173] A This distinct anisotropy has been experimentally confirmed via second harmonic generation measurements, which indicated that the magnetic order in CrSBr bilayers follows the anisotropic Heisenberg model, rather than the Ising or XY paradigms.[77] DFT studies have revealed that the triaxial magnetic anisotropy of CrSBr originates from the combined effects of MCA and MSA.[174, 175] A detailed investigation of Heisenberg exchange interactions and SIA by Wang et al. revealed that the exchange interactions are nearly isotropic, while the SOC of the ligand Br atom plays a dominant role in determining the SIA. In contrast, the SOC contributions from Cr and S atoms were found to be negligible when considered individually. Notably, when the SOC effects of both Br and Cr are taken into account simultaneously, the resulting SIA values closely match those observed experimentally, suggesting that the magnetic anisotropy is primarily governed by the strong SOC of the heavier ligand Br atoms, modulated by their hybridization with the Cr 3d states.

**3.2 Magnetic Anisotropy and Unquenched Orbital Magnetic Moments**

In magnetic materials, both magnetic anisotropy and unquenched orbital magnetic moments arise from SOC, and these two quantities are often interrelated. For instance, although VI$_3$ MLs and CrI$_3$ MLs are both 2D ferromagnets with honeycomb lattice coordinated by iodine atoms, the underlying electronic structures of their magnetic ions differ significantly. In CrI$_3$, the Cr$^{3+}$ ion adopts a 3$d^3$ configuration with a closed $t_{2g}^3$ shell in an octahedral crystal field, resulting in a high-spin state with S = 3/2 and negligible orbital angular momentum due to orbital quenching. In contrast, in VI$_3$, the V$^{3+}$ ion has a 3$d^2$ configuration with a partially filled $t_{2g}^2$ shell, yielding an S = 1 state and a greater susceptibility to unquenched orbital contributions. This difference in 3$d$ shells occupancy implies that the origin of magnetic anisotropy in VI$_3$ ML is different from that in CrI$_3$ ML, despite both systems exhibiting strong out-of-plane magnetic anisotropy. Using the DFT+U approach, Wang *et al*. reported that the out-of-plane MAE of VI$_3$ ML varies from 0.05 to 0.4 meV, depending on the value of the Hubbard $U$ parameter used.[177] The U-dependence arise from the fact that the band gap of VI$_3$ ML increases with increasing $U$, which in turn affects the SOC–induced splittings relevant to MAE.[177] In a separate study, Subhan *et al.* obtained has an out-of-plane MAE of 0.29 meV/cell for VI$_3$ ML.[178] Their atom-resolved analysis revealed that both vanadium atoms contribute equally to the anisotropy, with a MAE of approximately 0.15 meV per atom.[178] While both studies consistently predict an out-of-plane easy axis, the computed MAE values are significantly lower than the experimentally inferred value of about 1.16 meV, derived from helicity-resolved Raman spectroscopy measurements of the spin-wave gap.[179] Given that the open 3*d* shell of V$^{3+}$ ions, Yang *et al.* investigated the magnetic anisotropy of VI$_3$ ML by combining DFT calculations with crystal field level analyses.[180] Their results reveal that when the magnetization is oriented out-of-plane, the orbital magnetic moment per V ion can reach approximately 1.0 $\mu_B$/V, aligned antiparallel to the spin moment. In contrast, with in-plane magnetization, the orbital moment is significantly reduced to about 0.15 $\mu_B$ per V ion. As a result of such large orbital magnetic moment of V$^{3+}$ ions, the out-of-plane SIA and exchange anisotropy in VI$_3$ ML are calculated to be 15.9 and 0.67 meV, respectively.[180] Based on DFT calculations which employ the hybrid functional (HSE06) method of considering exact exchange, Zhao *et al.* obtained that the orbital magnetic moment in VI$_3$ ML is about 1.0 $\mu_B$/V for all magnetizations orientations.[181] This strong anisotropy in orbital magnetization leads to a pronounced out-of-plane SIA and exchange anisotropy, calculated to be 15.9 meV and 0.67 meV, respectively. [180] In a separate study, Zhao et al. [181] employed hybrid functional, which incorporates a portion of exact exchange, to examine the same system. Their calculations also predict an orbital magnetic moment of approximately 1.0 $\mu_B$/V per V ion; however, in contrast to Yang et al., this value is nearly independent of magnetization direction and their SIA is also low, 7.58 meV. The discrepancy is not unusual as the band splittings depend on the electron correlation and exchange effects.

It worth noting that although the MAEs reported by Yang[180] and Zhao[181] are significantly larger than the experimentally one,[178] their calculated orbital magnetic moments are in reasonable agreement with the experimental measurement of approximately 0.6 $\mu_B$/V.[182] In fact, two additional DFT calculations[183, 184] have also predicted the presence of large orbital magnetic moments in VI$_3$ ML, further reinforcing the idea that unquenched orbital angular momentum plays a central role in the magnetism of this material. Sandratskii *et al.*[184] further demonstrated that the anti-dimerization distortion of crystal structure significantly influences the orientation of the magnetic easy axis in bulk VI$_3$ [87]. On the other hand, VI$_3$ ML shares key structural characteristics with CrI$_3$ ML, including a honeycomb lattice and iodine ligand environment. Hence, the bond-dependent Kitaev interactions and off-diagonal symmetry exchange couplings Γ and Γ′, known to be relevant in CrI$_3$, could also play a non-negligible role in shaping the magnetic anisotropy of VI$_3$ ML.

The Ising-type 2D antiferromagnetism observed in FePS$_3$ ML has been closely linked to the presence of unquenched orbital magnetic moments on the Fe$^{2+}$ ions. Based on a comprehensive magnetic model incorporating an exceptionally large number of parameters derived from DFT calculations, Kim *et al.* pointed out the crucial role of orbital degrees of freedom in enhancing the magnetic anisotropy of FePS$_3$.[185] Supporting this picture, Amirabbasi *et al.* reported orbital ordering and a sizable orbital magnetic moment of about 0.8 $\mu_B$ per Fe atom in FePS$_3$ ML.[186] Further insights come from DFT calculations combined with crystal field level diagrams, which show that SOC induces a splitting of the half-filled minority-spin $e_g^\pi$ doublet in high-spin Fe$^{2+}$, resulting in an orbital magnetic moment of 0.76 $\mu_B$/Fe and FePS$_3$ ML and a remarkably large out-of-plane SIA energy of 19.4 meV.[187] These theoretical predictions are in good agreement with experimental findings: simulations of X-ray absorption spectroscopy based on ligand field multiplet theory reveal an even larger orbital magnetic moment of 1.02 ± 0.04 $\mu_B$/Fe and an exceptionally large out-of-plane MAE of 22 meV/Fe.[60] Physically, this giant MAE is attributed to the spin–orbit entangled nature of the $a_{1g}$ and $e_g^\pi$ states of the Fe$^{2+}$ ions (Figure 6D). Atomic orbital-resolved MCA calculations further confirm that the Fe atoms are the dominant contributors to the MAE in FePS$_3$ ML.[188, 189]

Magnetic anisotropies associated with unquenched orbital magnetic moments are also important in Kitaev magnetic materials, a class of spin-orbit assisted Mott insulators characterized by spin-orbit entangled $J_{eff} = 1/2$ states.[44] These $J_{eff} = 1/2$ states, characterized by substantial unquenched orbital magnetic moments, give rise to strongly anisotropic bond-dependent Kitaev and off-diagonal symmetric exchange interactions. As already mentioned above, α-RuCl$_3$ is a peculiarly compelling platform for investigating 2D quantum magnetism and proximity to a quantum spin liquid state. The pronounced magnetic anisotropy observed in α-RuCl$_3$ has been widely attributed to the presence of a sizable off-diagonal symmetric exchange interaction Γ. Early theoretical studies examining *g*-factor suggested that FM Kitaev interactions with relatively small off-diagonal symmetric exchange Γ and Γ′ could reproduce the observed magnetic anisotropy in α-RuCl$_3$. In contrast, for AFM Kitaev scenarios, a significantly larges and negative Γ coupling was required.[190] DFT calculations by Hou et al., which explicitly constrained the direction of orbital magnetic moments, further emphasize that the nearest neighbor off-diagonal symmetric exchange Γ plays a pivotal role in determining the preferred direction of magnetic moments in α-RuCl$_3$.[191] This theoretical insight was subsequently corroborated by resonant elastic

X-ray scattering experiments, which demonstrated that a large Γ along can account for the observed magnetic anisotropy in α-RuCl$_3$.[192] Physically, the off-diagonal symmetric exchange Γ effectively exerts a directional torque on the spin-orbital moments, energetically favoring spin alignment within the *ab* plane. This anisotropic exchange mechanism not only governs static magnetic anisotropy but also significantly influences the low-energy spin dynamics, thereby playing a central role in the complex magnetic behavior of α-RuCl$_3$.

It is valuable to compare magnetic anisotropies arising from ligand SOC with those driven by unquenched orbital magnetic moments on magnetic ions. Fundamentally, both mechanisms originate from SOC, and the distinction between them appears to be somewhat interpretive, serving primarily as a conceptual framework to aid physical understanding. Nevertheless, this classification proves useful when examining trends across different materials. In the case of CrX$_3$ compound, magnetic anisotropy is mainly related to ligand SOC. As a result, the MAE in CrX$_3$ can be turned across a wide range by halogen substitution, e.g., from out-of-plane MAE with iodine to in-plane MAE with chlorine. In contrast, the magnetic anisotropy in VI$_3$ is largely determined by the unquenched orbital magnetic moment on V$^{3-}$ ions. Consequently, substituting the ligand atoms in VI$_3$ with lighter halogens such as Br or Cl has limited impact on its magnetic anisotropy. This is supported by experimental observations, e.g., VBr$_3$ retains an out-of-plane magnetic moment alignment. [193] AFM VCl$_3$ ML grown on NbSe$_2$ substrate, which breaks both C$_3$ rotational and inversion symmetries, exhibits an *xz* magnetic easy plane,[194] differing from the xy plane seen in CrCl$_3$ ML.[76] Therefore, the division between ligand-driven and ion-driven anisotropy is useful especially when devising strategies to manipulate and enhance magnetic anisotropy in 2D vdW magnets through targeted chemical substitutions.

**3.3 Fe-based 2D itinerant ferromagnets**

Due to its relatively high Curie temperature, Fe$_3$GeTe$_2$ ML is a promising candidate for spintronic applications and has attracted significant research interest. As a metallic ferromagnet, its magnetization originates from a Stoner instability, primarily involving the $d_{xy}$ and $d_{x^2-y^2}$ orbitals of Fe atoms near the Fermi level.[103] DFT calculations by Zhuang *et al*. found that Fe atoms in Fe$_3$GeTe$_2$ ML have an average orbital magnetic moment of about 0.1 $\mu_B$[103] and the out-of-plane MCA energy is 0.92 meV/Fe. The in-plane MSA energy is negligible, only 36 μeV/Fe. Further insights into the origin of magnetic anisotropy were provided by Liu *et al.*, who analyzed the SOC matrix elements near the Fermi level. They showed that the positive (out-of-plane) contributions to MAE of Fe$_3$GeTe$_2$ mainly originate from SOC interaction across the unoccupied $d_{3z^2-r^2}$ orbitals and half-occupied $d_{xz}/d_{yz}$ orbitals of $Fe^{3+}$ ions. Conversely, the same orbital pairs in $Fe^{2+}$ ions give a negative (in-plane) contribution to MAE.[195] Their calculations yielded a total MAE of approximately 0.94 meV/Fe, consistent with previous DFT results.[103] Momentum-space resolved analysis by Park *et al.* demonstrated that the primary contributions to the out-of-plane MAE originate from electronic states along the K–H high-symmetry path in the Brillouin zone.[196]

As shown in Figure 7B, the strong SOC of ligand Te atoms was identified to play a dominant role in producing out-of-plane magnetic anisotropy in Fe$_3$GeTe$_2$. Wang *et al.* systematically examined the atom-specific SOC contributions to the MAE and found that the majority of the anisotropy arises from Te atoms, despite their weak induced magnetization from neighboring Fe atoms.[197] Notably, turning off the SOC on Fe atoms has a negligible impact on the MAE, while eliminating SOC on Te atoms dramatically reduces the MAE to 0.2 meV/Fe, underscoring their dominant role.[198] Furthermore, the strong contribution from Te is attributed to SOC-induced mixing between its $p_x$ and $p_y$ orbitals near the Fermi level, as well as to the presence of a topological nodal point just below the Fermi level at the K points in the Brillouin zone.[199]

Experimentally, a field-dependent anisotropic magnetic coupling has been observed in Fe$_{3-x}$GeTe$_2$ ($x \approx 0.28$),[200] suggesting a possible connection between the material's magnetic anisotropy and anisotropic exchange interactions. DFT calculations for Fe$_3$GeTe$_2$ have revealed that the nearest-neighbor exchange anisotropy, $\Delta J = J^x - J^z$, is approximately 0.2 meV/$\mu_B^2$, which is an order of magnitude larger than the SIA (0.02 meV/$\mu_B^2$).[118]

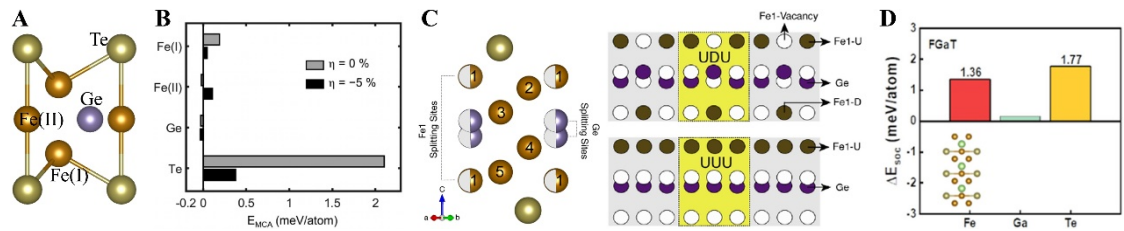

*Figure 7.* A) *Illustration of the label of atoms for Fe$_3$GeTe$_2$. B) The atom-decomposed MCA energy (E$_{MCA}$) for Fe$_3$GeTe$_2$ ML (corresponding to $\eta$ = 0%). Reproduced with permission.*[198] *Copyright 2024, American Physical Society. C) Right panel: the side view of Fe$_5$GeTe$_2$ ML with Fe1-Ge split sites. Brown, purple, and dark green balls show Fe, Ge, and Te atoms, respectively. Left panel: schematic representation of UDU and UUU configurations in upper and lower panels, respectively. Circles filled with color and white show the presence and absence of atoms, respectively. Reproduced with*

*permission.[201] Copyright 2022, The Authors. D) The atom-resolved $\Delta E_{SOC}$ of Fe$_3$GaTe$_2$. Reproduced with permission.[202] Copyright 2024, American Chemical Society.*

Fe$_3$GeTe$_2$ exhibits an out-of-plane magnetic anisotropy that is about a three times larger than that of Fe$_4$GeTe$_2$.[105] To elucidate the electronic origins underlying this stark contrast in magnetic anisotropy, chemical bonding and Coulomb screening of Fe$_3$GeTe$_2$ and Fe$_4$GeTe$_2$ are examined by X-ray absorption spectroscopy and DFT calculations.[203] The analysis revealed that the Fe dumbbell structures in Fe$_3$GeTe$_2$ result in $d_{3z^2-r^2}$ bonding and antibonding states, which are effectively split from the $t_{2g}$ manifold. In contrast, Fe atoms in Fe$_4$GeTe$_2$ form bonds among themselves with more symmetric electronic structure. As a result, the Coulomb interaction is more effectively screened in Fe$_4$GeTe$_2$ than Fe$_3$GeTe$_2$, according to X-ray absorption spectroscopy and DFT data.[203]

Unlike Fe$_3$GeTe$_2$ and Fe$_4$GeTe$_2$, Fe$_5$GeTe$_2$ has a unique structural feature: the Fe1 atoms have two equally probable sites with partial occupancy.[106, 112, 204, 205] As a result, Ge sites are also split (see the right panel in Figure 7C).[106, 112, 116] A high-brilliance laboratory X-ray experiment further revealed that Fe$_5$GeTe$_2$ has a long-range ordered $\sqrt{3} \times \sqrt{3}$-R30º superstructure, a minority phase associated with ordered Te vacancies. This superstructure is inherently two-dimensional, lacking lattice periodicity perpendicular to the vdW layers.[206] Ershadrad *et al.* modeled Fe$_5$GeTe$_2$ ML using two representative Fe1 configurations (see the left panel in Figure 7C).[201] Their DFT calculations showed that Fe$_5$GeTe$_2$ ML has a weak out-of-plane MAE of 0.021 meV/Fe in a UDU configuration. For the UUU configuration, they obtained an in-plane MAE of 0.112 meV/Fe. These results are consistent with the sample-dependent magnetic anisotropies observed experimentally in various studies..[111-115] Using DUU configuration, which is considered more energetically favorable, [207] Hu *et al.* also found an in-plane magnetic anisotropy in Fe$_5$GeTe$_2$ ML based on DFT calculations.[208] Notably, they attributed a major role to the Te's SOC. On the experimental side, a comparison between measured critical exponents and theoretical predictions from various magnetic exchange models suggests an enhancement of anisotropic exchange below the Curie temperature in Fe$_5$GeTe$_2$.[209] However, broadband FM resonance spectroscopy of bulk single-crystal near $T_C$ indicates that anisotropic symmetric exchange is negligible, if present at all.[210] X-ray and electron microscopy studies by Birch *et al.* suggested that the ordering of the Fe1 sublattice may alter the out-of-plane magnetic anisotropy in Fe$_5$GeTe$_2$.[211] To fully elucidate the origins and behavior of magnetic anisotropy in Fe$_5$GeTe$_2$, additional in-depth experimental and theoretical studies are still needed.

DFT studies have shown that $Fe_3GaTe_2$ ML, which is isostructural to $Fe_3GeTe_2$, has an out-of-plane MAE of 0.31 meV/Fe.[212] An atom-resolved analysis further revealed that the out-of-plane SIA is $Fe_3GaTe_2$ ML reaches 0.69 meV/Fe, with dominant contributions from Fe and Te atoms (Figure 7D).[202] However, a separate DFT study by Xi *et al.* reported that the nearest neighbor exchange anisotropy $\Delta J = J^x - J^z$ in $Fe_3GaTe_2$ is as large as 3.2 meV/$\mu_B^2$, which is much larger than the rescaled SIA (0.05 meV/$\mu_B^2$).[118] This contrast suggests that the strong out-of-plane magnetic anisotropy in $Fe_3GaTe_2$ originates predominantly from Ising-type exchange anisotropy, rather than from SIA. This result suggested that the strong out-of-plane magnetic antitropy of $Fe_3GaTe_2$ originates almost from the strong Ising-type exchange anisotropy. Supporting this theoretical insight, critical exponent measurements indicate that $Fe_3GaTe_2$ behaves as a vdW ferromagnet governed by a 3D Ising model with long-range interactions.[213]

**4. Impact of Magnetic Anisotropy on Electronic Properties in 2D vdW Magnets**

Beyond determining the magnetic orientation and phase transitions of 2D vdW magnets, magnetic anisotropy serves as a powerful tuning parameter for modulating their electronic and topological properties. By constructing 2D vdW heterostructures that combine magnetic layers with other functional materials, novel quantum phenomena, such as the quantum anomalous Hall effect, axion insulator phases, and valley polarization, can be realized through the magnetic proximity effect. Moreover, the properties of 2D vdW magnetic heterostructures can be further engineered through layer stacking, elemental substitution, doping, and strain application, offering rich opportunities to achieve diverse topological spin textures and enhanced Curie temperatures, both critical for the development of practical spintronic, valleytronic, topotronic, and quantum devices. As discussed below, magnetic anisotropy plays a pivotal role in enabling and controlling these emergent phenomena in 2D vdW magnetic heterostructures.

**4.1 Spin-orientation-dependent electronic properties in 2D vdW magnets**

Since the SOC Hamiltonian term $\xi \boldsymbol{L} \cdot \boldsymbol{S}$ depends on the spin orientation, the electronic band structure may display a change when the spin orientation is adjusted by an external magnetic field. This phenomenon is called a magneto-band-structure (MB) effect.[13] Such effect is illustrated by a 2D three-*p*-orbital ($p_x$, $p_y$, and $p_z$) toy model as shown in Figure 8A.[13] Interestingly, $CrI_3$ ML is demonstrated by a DFT calculation study to show a giant MB effect (Figure 8B).[13] When switching its spin magnetic moment from out-of-plane to in-plane, $CrI_3$ ML experiences a direct-to-indirect bandgap transition which can be observed by measuring the photoluminescence.

Additionally, doped CrI$_3$ ML shows a significantly changed Fermi surface with different spin magnetic moment orientations (Figure 8C), which can achieve giant anisotropic magnetoresistance.[13] Finally, an novel topological phase transition from a single-spin Dirac semimetal to a Chern topological insulator can be realized in electron doped CrI$_3$ ML.[13]

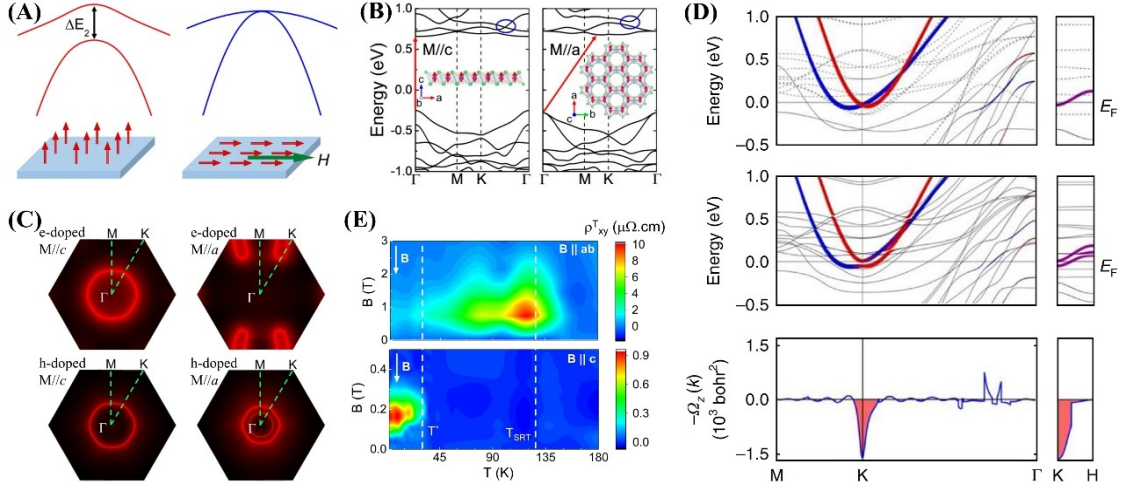

*Figure 8.* A) Schematic of magneto-band-structure effect. Left and right panels show the calculated band splitting using a toy model for a 2D magnet with spin orientations along out-of-plane and in-plane directions, respectively. B) Band structure of CrI$_3$ ML with different spin orientations. C) Spin orientation dependent Fermi surfaces of CrI$_3$ ML doped with 0.2 electrons or holes. A), B) and C) are reproduced with permission.[13] Copyright 2018, American Chemical Society. D) Band structures of Fe$_3$GeTe$_2$ bulk. Upper panel: spin-polarized band structure without SOC in DFT calculations. Middle panel: band structure with SOC included in DFT calculations when the spin orientation is out-of-plane. Bottom panel: the corresponding Berry curvature to the middle panel. Reproduced with permission.[214] Copyright 2018, The Author(s). E) Contour plots of topological Hall effect contributions in Fe$_4$GeTe$_2$ are presented for B||ab and B||c as a function of magnetic field and temperature. Reproduced with permission.[215] Copyright 2025, Arxiv.

Similar to CrI$_3$ ML, spin orientations play a vital role in determining the topological properties of Fe$_3$GeTe$_2$. As shown in Figure 7D, a nodal point appears below the Fermi level at the *K* point in the spin-polarized band structure of bulk Fe$_3$GeTe$_2$, as calculated by DFT. [214] This nodal point extends along the *K-H* line, forming a topological nodal line protected by $C_{3z}$ and $\tilde{C}_{6z}M_y$ symmetries.[214] When SOC is included and the spin orientation is set to out-of-plane in DFT calculations, this topological nodal line is gapped, leading to emergence of large Berry curvature near the gapped region (middle and bottom panels in Figure 8D). Experimentally, this gapping manifests as a large anomalous Hall current observed in $Fe_3GeTe_2$ crystals, providing evidence for the topological nature of the electronic states.[214] In contrast, when the spin orientation is in-plane, the nodal line remains gapless, preserving the topological line degeneracy. [199]

$Fe_4GeTe_2$ exhibits a temperature-driven spin reorientation transition,[100, 105, 108, 109] therefore, it is a natural playground to explore the effect of spin orientations on magneto-transport and topological properties. By studying various transport properties like the ordinary Hall effect, resistivity, magnetoresistance and the anomalous Hall conductivity, it is suggested that the resistivity of a $Fe_4GeTe_2$ nanoflake is governed by the electron-electron scattering, inelastic scattering contributions, electron-magnon and electron-phonon when it has an out-of-plane spin orientation.[15] When the $Fe_4GeTe_2$ nanoflake has an in-plane spin orientation, its resistivity is dominated by the electron-phonon scattering.[15] Interestingly, it is even reported that, in a $Fe_4GeTe_2/Fe_4GeTe_2$ homojunction, the temperature-dependent magnetoresistance is negative when its spin orientation is in-plane but positive when its spin orientation is out-of-plane.[14] Besides, Bera *et al*. observed an anomalous Hall effect which is induced by Berry curvatures in $Fe_4GeTe_2$ when its spin orientation is along the out-of-plane magnetic easy axis.[216] They further showed based on DFT calculations that the anomalous Hall conductivity of $Fe_4GeTe_2$ decreases with its spin orientation being rotated from the out-of-plane *z*-axis towards the in-plane *x*-axis, eventually reaching zero when its spin orientation is along the *x*-axis. All these findings indicate that the magneto-transport and topological properties of $Fe_4GeTe_2$ are dependent on its flexible spin orientations, which lies an important foundation for its potential to explore novel spintronic devices.

Another important consequence of tuning magnetic anisotropy is the ability to stabilize and manipulate complex spin textures in real space, such as magnetic skyrmions, chiral domain walls, spin spirals, and merons, in 2D vdW magnetic materials. The emergence and stability of these textures depend sensitively on the interplay among Heisenberg exchange interactions, magnetic anisotropy, Dzyaloshinskii–Moriya interactions, and Zeeman energy.[9] Angle-dependent magneto-transport measurements on $Fe_4GeTe_2$ single crystals[215] have revealed that a large topological Hall effect in multilayer samples, attributed to spin reorientation transitions that give rise to non-coplanar spin textures. As shown in Figure 8E, this topological Hall effect persists throughout the entire temperature range from $T_{SRT}$ down to the lowest measured temperature when the magnetic field is applied within the *ab*-plane (B||*ab*). In contrast, when the field is aligned along the *c*-axis (B||*c*), the topological Hall effect appears only within a narrower temperature window below 30 K. Of particular interest is the experimental observation of coexistence of merons and magnetic skyrmions in $Fe_{5-x}GeTe_2$, as demonstrated by Casas *et al*. [217] This behavior arises from the weak and Fe-deficiency-dependent MCA in $Fe_{5-x}GeTe_2$. In regions with out-of-plane anisotropy, skyrmions are stabilized, whereas merons are found in domains characterized by in-plane anisotropy.[217]

**4.2 Heterostructures of Topological and 2D vdW magnetic Materials**

Inducing long-range spin polarization into topological materials is an attractive strategy for realizing exotic quantum phases, such as QAHE and axion insulator state, which host robust, dissipationless edge currents and topologically quantized magnetoelectric coupling.[16] While conventional approaches include doping topological materials with 3d transition metal elements [218] or interfacing them with non-vdW magnetic materials,[219] a particularly appealing alternative is the construction of heterostructures combining 2D vdW magnetic materials with topological insulators. This approach leverages the mechanical flexibility, clean interfaces, and tunability of vdW heterostructures, offering a versatile platform for exploring proximity-induced topological magnetism and quantum transport phenomena.[220]

By co-depositing Mn and Se onto the surface of the 3D TI, $Bi_2Se_3$, Hirahara *et al*. fabricated a self-assembled $MnBi_2Se_4/Bi_2Se_3$ heterostructure (Figure 9A).[221] Their experimental studies revealed that this heterostructure exhibits robust ferromagnetism with out-of-plane magnetic anisotropy, persisting up to room temperature. More remarkably, ARPES measurements showed a clear Dirac gap opening of approximately 100 meV, while the rest of the band structure remains largely intact (Figures 9B–9C). Complementary DFT calculations confirmed that $MnBi_2Se_4/Bi_2Se_3/MnBi_2Se_4$ possesses a nontrivial Chern number of $C_N = -1$, identifying the system as a quantum anomalous Hall phase.

Using MBE, Rienks *et al*. reported the formation of a vdW $MnBi_2Te_4/Bi_2Te_3$ heterostructure by growing Mn-doped $Bi_2Te_3$ on suitable substrates. This heterostructure features a self-organized alternating sequence of $MnBi_2Te_4$ septuple layers (SLs) and $Bi_2Te_3$ quintuple layers (QLs) (Figure 9D).[222] Interestingly, similar layered heterostructures were also observed in Mn-doped $Bi_2Se_3$ and Mn-doped $Sb_2Te_3$, resulting in $MnBi_2Se_4/Bi_2Se_3$ and $MnSb_2Te_4/Sb_2Te_3$, respectively. DFT calculations indicated that both $MnBi_2Te_4/Bi_2Te_3$ and $MnSb_2Te_4/Sb_2Te_3$ structures have out-of-plane magnetic anisotropy, whereas $MnBi_2Se_4/Bi_2Se_3$ has an in-plane anisotropy due to its relatively weaker SOC. As a result, ARPES measurements showed that $MnBi_2Te_4/Bi_2Te_3$ hosts a sizable magnetic gap of $90\pm10$ meV at the Dirac point (Figure 9E-9G), while $MnBi_2Se_4/Bi_2Se_3$ only exhibits a nonmagnetic gap at its Dirac point. This contrasting behavior underscores the critical role of magnetic anisotropy in realizing nontrivial band topology in heterostructures that combine 2D vdW magnetic materials with 3D TIs. Since MnSb$_2$Te$_4$/Sb$_2$Te$_3$ is p-type with out-of-plane magnetic anisotropy and a magnetic Dirac gap while MnBi$_2$Te$_4$/Bi$_2$Te$_3$ is n-type, alloying to form Mn(Bi,Sb)$_2$Te$_4$/Bi$_2$Te$_3$ enables band and anisotropy engineering within a single heterostructure, offering an appealing route toward achieving robust edge-state transport in QAHE devices.

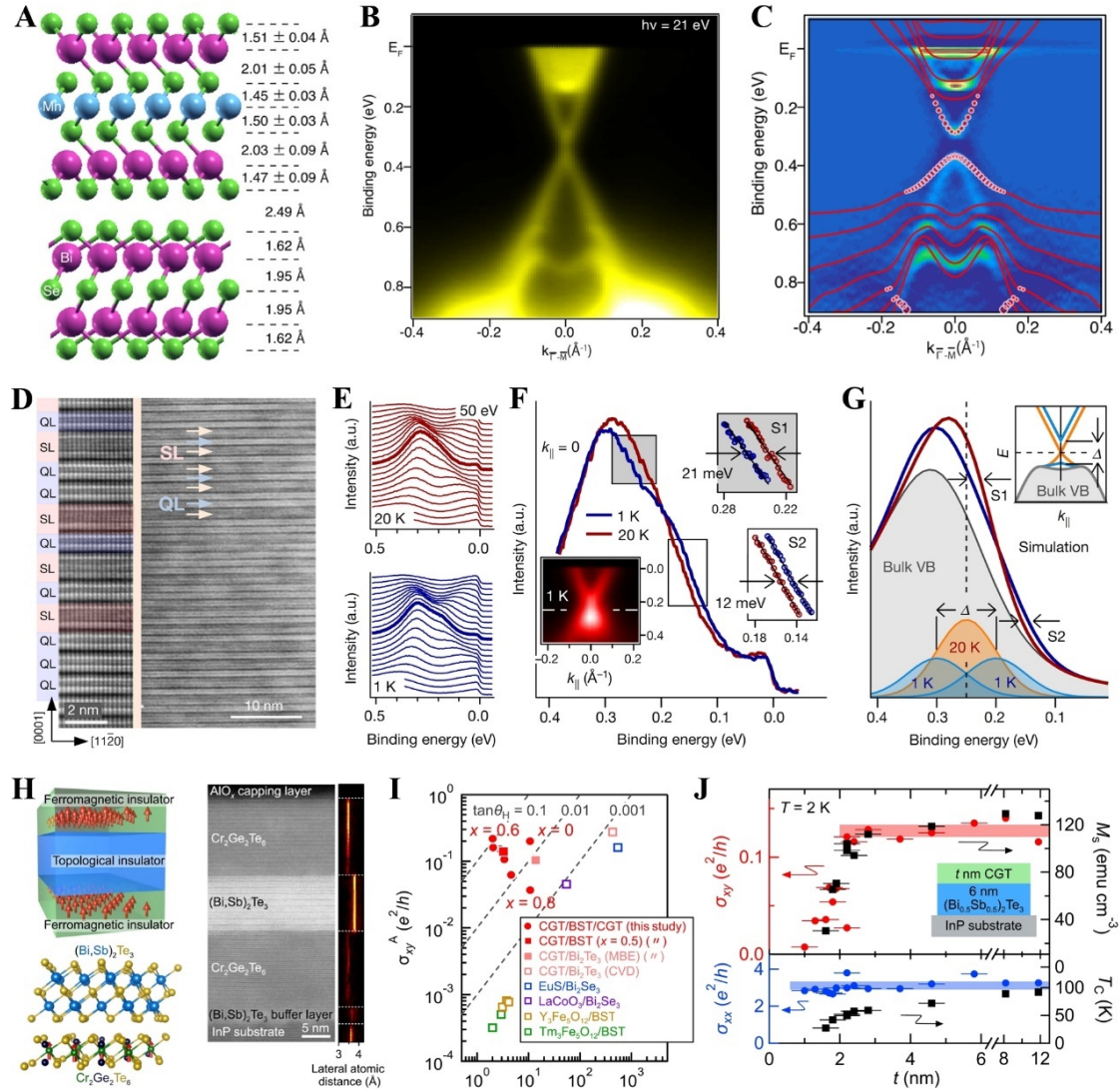

*Figure 9.* A) *The side view of the optimized crystal structure of MnBi$_2$Se$_4$/Bi$_2$Se$_3$ heterostructure. B) The band dispersion of MnBi$_2$Se$_4$/Bi$_2$Se$_3$ heterostructure measured along the $\bar{\Gamma} - \bar{M}$ direction taken at hv = 21 eV. C) The calculated band dispersion overlapped with the experimental data. Reproduced with the permission.[221] Copyright 2017, American Chemical Society. D) High-resolution scanning transmission electron microscopy cross-sections of Mn-doped Bi$_2$Te$_3$. E), F) and G) show the ARPES for Bi$_2$Te$_3$ with 6% Mn, above and below the $T_C$ of around 10 K. The spectra in F) and G) and those marked by thick lines in E) correspond to the center of the surface Brillouin zone at $k_{||}$ = 0. D), E), F) and G) are reproduced with permission.[222] Copyright 2019,*

*The Author(s), under exclusive license to Springer Nature Limited.* H) *Right panel: the schematic of the interfacial exchange coupling in a heterostructure of 3D TI thin film and 2D FM insulator and the crystal structures of TI (Bi,Sb)$_2$Te$_3$ and FM insulator Cr$_2$Ge$_2$Te$_6$. Left panel: the cross-sectional image of Cr$_2$Ge$_2$Te$_6$/(Bi,Sb)$_2$Te$_3$/Cr$_2$Ge$_2$Te$_6$ heterostructure.* I) *The anomalous Hall conductivity ($\sigma_{xy}^A$) is plotted against $\sigma_{xx}$ for Cr$_2$Ge$_2$Te$_6$/(Bi,Sb)$_2$Te$_3$/Cr$_2$Ge$_2$Te$_6$ and other FM insulator-TI heterostructures.* J) *Upper panel: Cr$_2$Ge$_2$Te$_6$ thickness (t) dependence of $\sigma_{xy}$ and M$_s$ at 2 K under zero magnetic field. Bottom panel: same as the upper panel but for $\sigma_{xx}$ and T$_C$. The inset shows the schematic of (Bi$_{0.5}$Sb$_{0.5}$)$_2$Te$_3$/Cr$_2$Ge$_2$Te$_6$ bilayer structure. H), I) and J) are reproduced with permission.*[223] *Copyright 2019, American Physical Society.*

A large anomalous Hall effect has been experimentally observed in a heterostructure composed of 2D vdW FM insulator, Cr$_2$Ge$_2$Te$_6$, and a thin film of 3D TI, (Bi,Sb)$_2$Te$_3$.[223] Using MBE, Mogi *et al.* fabricated a Cr$_2$Ge$_2$Te$_6$/(Bi,Sb)$_2$Te$_3$/Cr$_2$Ge$_2$Te$_6$ heterostructure (Figure 9H)[223] and found that it exhibits out-of-plane ferromagnetism with a anomalous Hall conductivity as high as 0.2 $e^2/h$ (Figure 9I), despite the weak magnetization in the TI layers. This indicates that out-of-plane magnetization of Cr$_2$Ge$_2$Te$_6$ can effectively magnetize the topological surface states of (Bi,Sb)$_2$Te$_3$ through the magnetic proximity effect, thereby opening a sizable exchange gap and inducing anomalous Hall conductivity (Figure 9J). The Cr$_2$Ge$_2$Te$_6$/graphene heterostructure has emerged as a promising platform for exploring nontrivial topological phases and magnetic proximity effects, both theoretically and experimentally. Using DFT calculations, Zhang *et al.* predicted that an out-of-plane magnetization induced QAHE in this vdW heterostructure.[224] Importantly, their calculations showed that the QAHE is robust against variations in the stacking configuration between graphene and Cr$_2$Ge$_2$Te$_6$, and that the Fermi level lies precisely within the nontrivial band gap, ensuring the quantized Hall conductance. Experimentally, Yao *et al.*[225] also observed the anomalous Hall effect in Cr$_2$Ge$_2$Te$_6$/graphene. Complementary spin precession measurements by Yang *et al.* revealed that graphene in this heterostructure acquires appreciable spin polarization and SOC from Cr$_2$Ge$_2$Te$_6$,[226] enabling the anomalous Hall effect for the use in seamless, gate-tunable spintronic devices.

By combining DFT calculations with a four-band model Hamiltonian simulations, Hou *et al.* investigated the potential for realizing QAHE in vdW heterostructure composed of Bi$_2$Se$_3$ and CrI$_3$ (Figure 10A).[227] They demonstrated that out-of-plane magnetization of the FM CrI$_3$ ML effectively magnetizes the topological surface states of Bi$_2$Se$_3$ films and creates a sizable spin splitting at the Dirac point (Figure 10B). By varying the thickness of 3D TI $Bi_2Se_3$ films, they found that the $CrI_3/Bi_2Se_3/CrI_3$ heterostructure can support QAHE at a temperature of a few tens of Kelvins when the $Bi_2Se_3$ film exceeds 5 QLs in thickness (Figure 10C). By embedding a $CrI_3$ ML within $Bi_2Se_3$, Chen *et al*. showed that the topologically nontrivial gap can be enlarged, up to 30 meV, in the multilayer structure $Bi_2Se_3/CrI_3/Bi_2Se_3/CrI_3/Bi_2Se_3$.[228] It is worth noting that further increasing the thickness of $CrI_3$ in $CrI_3/Bi_2Se_3$ heterostructures is unlikely to enhance the proximity-induced Dirac gap. This limitation arises from two key factors: (1) the spatial extent of the topological surface states in $Bi_2Se_3$ is inherently confined, restricting effective exchange coupling to only the nearest $CrI_3$ layers; and (2) $CrI_3$ exhibits AFM interlayer ordering in its multilayer form.[5] In a comparable case, it has been shown than shown that the FM $MnBi_2Se_4$ ML produces a larger Dirac gap in the $MnBi_2Se_4/Bi_2Se_3$ heterostructure than the AFM $Mn_2Bi_2Se_5$ layer[229, 230] Additionally, Zhang *et al*. demonstrated that the FM $CrI_3$ ML can induce a very large magnetic exchange field (about 150 meV) in graphene within a $CrI_3$/graphene heterostructure, providing that the $CrI_3$-graphene interlayer distance is reduced from 3.3 Å to 2.4 Å through external compression[231].

The vdW heterostructure of 2D TI Bi bilayer and FM $MnBi_2Te_4$ ML is of high interest for examining the close relationship between magnetic anisotropy and QAHE.[11] Note that the lattice constants of Bi bilayer and $MnBi_2Te_4$ ML are well matched which is advantageous for experimental synthesis of Bi/$MnBi_2Te_4$ vdW heterostructure (Figure 10D). DFT calculations show that Bi/$MnBi_2Te_4$ has an in-plane magnetic easy, i.e., in-plane magnetic anisotropy and the energy barrier of spin rotation within the *x-y* plane is negligibly small. Considering the in-plane magnetic anisotropy of Bi/$MnBi_2Te_4$, band structures and edges state study indicated that this heterostructure can realize an-plane magnetization induced QAHE with a Chern number of $C_N = +1$ in this heterostructure when aligning its magnetization along the *x* axis (Figure 10E). When its magnetization is rotated by $\Phi = 60°$ with respect to the *x* axis, Bi/$MnBi_2Te_4$ exhibits a QAHE with Chern number of $C_N = +1$. As a result of the spin rotation within the *x-y* plane, the Chern number of Bi/$MnBi_2Te_4$ oscillates between +1 and -1 with an interval of 60° (Figure 10F), consistent with the crystal symmetry of Bi/$MnBi_2Te_4$. When its magnetization is aligned along the out-of-plane direction which can be realized by applying an external magnetic field, strain, or twisting, Bi/$MnBi_2Te_4$ surprisingly displays a QAHE with high Chern numbers of $C_N = \pm 3$. Therefore, the QAHE in Bi/$MnBi_2Te_4$ is dictated unambiguously by its magnetic anisotropy. Note that an in-plane magnetization induced QAHE is also theoretically predicted in other 2D vdW magnetic materials, such as LaCl, $NiAsO_3$ and $PbSbO_3$ MLs.[232, 233]

An axion insulator phase in vdW heterostructures of 2D magnetic insulators and 3D

TI thin films is also a topological quantum state which arises from the interplay between the out-of-plane magnetization of the former and the topological surface state of the latter. It has been experimentally reported in conventional FM/TI/FM heterostructures whose two FM materials has different coercive fields ($H_c$).[234, 235] Given the narrow external magnetic field range for the appearance of the axion insulator phase in the FM/TI/FM heterostructures, a FM/TI/AFM heterostructure of 3D TI thin films and 2D vdW magnetic materials is put forward to realize the axion insulator phase in a wide external magnetic field range (Figure 10H). The core of realizing axion insulator phase in such heterostructure is still magnetizing the topological surface states of 3D TI thin films with the out-of-plane magnetizations of 2D vdW FM and AFM materials. Inspired by the experimentally synthesized FM $MnBi_2Se_4$,[221] the AFM $Mn_2Bi_2Se_5$ ML with an out-of-plane magnetic anisotropy is predicted and $MBi_2Se_4/Bi_2Se_3/Mn_2Bi_2Se_5$ is theoretically proposed to achieve the axion insulator phase.[230] The DFT calculated band structures and surface-projected Berry curvatures (Figure 10I-10K) indicate that the top and bottom surfaces of $MBi_2Se_4/Bi_2Se_3/Mn_2Bi_2Se_5$ have opposite half-quantum Hall conductance, i.e., $\sigma_{xy}^t = e^2/2h$ and $\sigma_{xy}^b = -e^2/2h$. So, FM/TI/AFM heterostructures with out-of-plane magnetic anisotropy are feasible to realize the axion insulator.

Ferromagnetically ordered heterostructures of 2D vdW magnetic materials and 3D TI thin films is also suggested to achieve the axion insulator phase in a high magnetic field, which facilitates the experimental detections of the topological magnetoelectric effect.[236] The underlying mechanism for the existence of the axion insulator phase in such heterostructures is that the out-of-plane FM magnetizations of the two different 2D vdW magnetic materials introduce opposite exchange fields into the topological surface state of 3D TI thin films. As a result of the opposite exchange fields, such heterostructure can give rise to the opposite half-quantum Hall conductance at its two surfaces, namely, the axion insulator state, in a ferromagnetically ordered magnetization. A theoretical investigation based on DFT calculations and a four-band model suggested that the ferromagnetically order heterostructure $CrI_3/Bi_2Se_3/MnBi_2Se_4$ indeed hosts the axion insulator phase.[236]

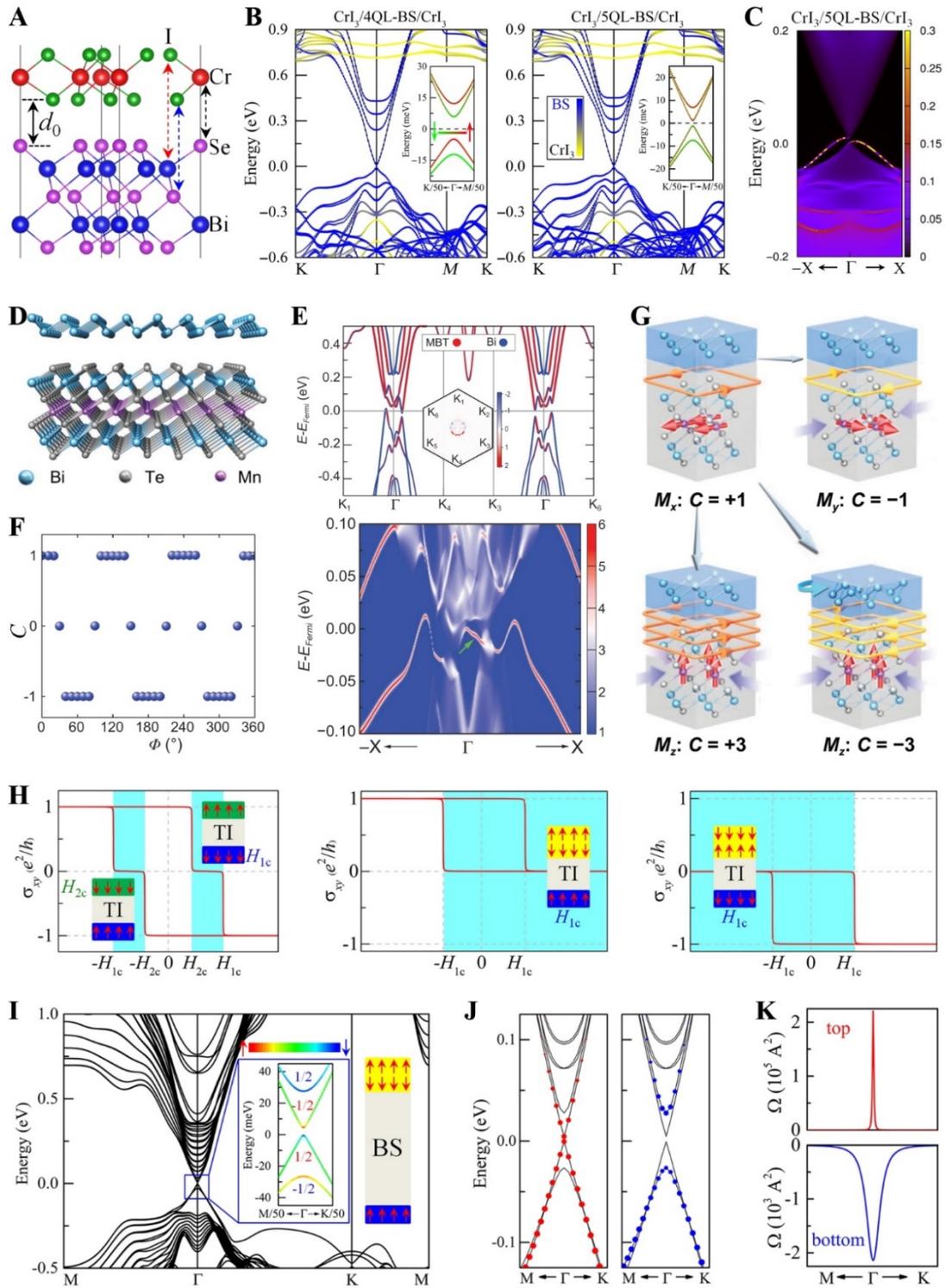

*Figure 10.* A) Stacking configuration at the interface between CrI$_3$ ML and Bi$_2$Se$_3$ thin films. B) DFT calculated band structures of CrI$_3$/4QL-BS/CrI$_3$ and CrI$_3$/5QL-BS/CrI$_3$ heterostructures. C) The chiral edge state of the Chern insulator CrI$_3$/5QL-BS/CrI$_3$ ribbon. A), B) and C) are reproduced with permission.[227] Copyright 2019, The Authors, some rights reserved; exclusive licensee AAAS. D) The crystal structure of Bi/MnBi$_2$Te$_4$ heterostructure. E) DFT calculated band structure (upper panel) and the edge states highlighted by green arrows (bottom panel) for Bi/MnBi$_2$Te$_4$ with an in-plane

*magnetization along Φ = 0°. F) The Chern number as a function of the in-plane magnetization direction in Bi/MnBi$_2$Te$_4$, where Φ is the angle between magnetization and x direction. G) Schematic diagram shows magnetic anisotropy controlled topological phases in Bi/MnBi$_2$Te$_4$. Strain is denoted by purple arrows. Reproduced with permission.[11] Copyright 2023, The Author(s) 2023. H) A comparison of the presence of axion insulator phase in FM/TI/FM and FM/TI/AFM heterostructures. The light cyan highlights the external magnetic field ranges that give rise to axion insulator phase. FM and AFM insulators are sketched by blue and yellow blocks, respectively. The insets show the magnetization configurations of axion insulator phase. I) DFT band structure of the axion insulator phase in MBi$_2$Se$_4$/Bi$_2$Se$_3$/Mn$_2$Bi$_2$Se$_5$. J) Top-QL-Bi$_2$Se$_3$ (right panel) and bottom-QL-Bi$_2$Se$_3$ (left panel) projected bands in MBi$_2$Se$_4$/Bi$_2$Se$_3$/Mn$_2$Bi$_2$Se$_5$. K) The Berry curvature (Ω) of the occupied topological surface states of top (upper panel) and bottom (bottom panel) surfaces in MBi$_2$Se$_4$/Bi$_2$Se$_3$/Mn$_2$Bi$_2$Se$_5$. Reproduced with permission.[230] Copyright 2019, American Chemical Society.*

**4.3 Valley Physics and Magnetic Skyrmions in 2D vdW Magnetic Heterostructures**

Valley semiconductors, such as MoS$_2$ ML, have valley pseudospin degrees of freedom in their band structures and valleytronics applications.[237] These valleys are optically addressable but degenerate in energy, so lifting their energy degeneracy by breaking time-reversal symmetry is crucial to valley manipulations. Usually, valley splitting in valley semiconductors can be induced by an out-of-plane magnetization through the magnetic proximity effect. By growing heterostructures of the ultrathin CrI$_3$ and WSe$_2$ ML (Figure 11A), Zhong *et al*. observed a valley splitting of about 3 meV at a zero applied magnetic field below the $T_C$ of CrI$_3$, comparable with the DFT calculated valley splitting of 2 meV.[72, 238] Besides, the conduction band of WSe$_2$ ML lies above the spin-polarized $e_g$ band of CrI$_3$ in CrI$_3$/WSe$_2$, which leads to a valley polarization (Figure 11B). Through the optical control of CrI$_3$ magnetization, a wide continuous tuning of the valley splitting and valley polarization is realized in CrI$_3$/WSe$_2$.[239] Based on DFT calculation, Zhang *et al*. demonstrated that the interfacial atom superposition plays an important role and a W-Cr superposition is essential for a relatively large valley splitting in CrI$_3$/WSe$_2$.[240] They also demonstrated that valley splitting can be tuned in the trilayer CrI$_3$/WSe$_2$/CrI$_3$ from nearly zero to more than two times of that in WSe$_2$/CrI$_3$ by manipulating the layer alignment. Surprisingly, it is theoretically suggested that the valley splitting in the twisted WSe$_2$/CrI$_3$ can be strongly enhanced compared with the non-twisted WSe$_2$/CrI$_3$.[241, 242]

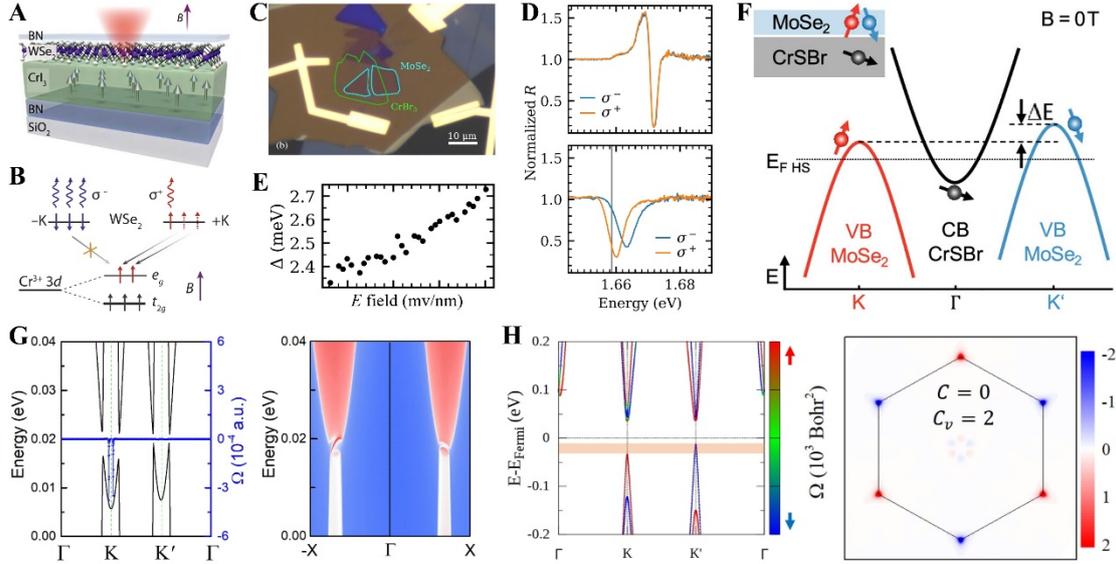

*Figure 11.* A) Schematic of vdW heterostructure CrI$_3$/WSe$_2$ encapsulated by h-BN. B) Schematic depicts the spin orientation-dependent charge hopping between WSe$_2$ and CrI$_3$. A) and B) are reproduced with permission.[72] Copyright 2017, The Authors, some rights reserved. C) Optical micrograph of CrBr$_3$/MoSe$_2$ heterostructure with MoSe$_2$ and CrBr$_3$ outlined in blue and green, respectively. D) Reflection spectra of bare MoSe$_2$ ML (upper panel) and CrBr$_3$/MoSe$_2$ (bottom panel). E) Dependence of the valley splitting on the electric field E. C), D) and E) are reproduced with permission.[243] Copyright 2020, American Physical Society. F) Schematic of band structure and valley splitting in MoSe$_2$/CrSBr. VB and CB stand for valence band and conduction band, respectively. Copyright 2024, The Authors. G) Left panel: band structure and Berry curvature of WSe$_2$/CrBr$_3$. Right panel: edge state of a half-infinite WSe$_2$/CrBr$_3$ heterostructure. Reproduced with permission.[244] Copyright 2020, American Physical Society. H) The out-of-plane spin-projected band structure (left panel) and the distribution of Berry curvature ($\Omega$) (right panel) of VBi$_2$Te$_4$/Ge with SOC being included. Reproduced with permission.[245] Copyright 2024, American Physical Society.

An out-of-plane magnetization induced valley splitting is observed in the heterostructure composed of of CrBr$_3$ bilayer and MoSe$_2$ ML. Figure 11C shows the optical micrograph of an experimentally grown CrBr$_3$/MoSe$_2$ heterostructure in a resonant optical spectroscopy measurement.[243] In the absence of an external magnetic field, the $K$ and $K'$ valley excitons are degenerate in bare MoSe$_2$ and its spectrum has no polarization dependence (the upper panel in Figure 11D). In contrast, a valley splitting of 2.9 meV emerges in the CrBr$_3$/MoSe$_2$ heterostructure region (the bottom panel in Figure 11D). This valley splitting is further demonstrated to be directly linked to the magnetization of CrBr$_3$. In addition, the valley splitting in CrBr$_3$/MoSe$_2$ heterostructure exhibits an approximately linear dependence on the applied out-of-plane electric field (Figure 11E), which is in line with theoretical results[246] and could has implications for gate-tunable valleytronic devices.

Although a CrSBr film is a layered A-type vdW antiferromagnet with an in-plane magnetic anisotropy, it can exert a magnetic proximity interaction on the valleys. By investigating the magneto photoluminescence of MoSe$_2$ ML on the CrSBr film, Brito *et al*. observed a clear influence of the CrSBr magnetic order on the optical properties of MoSe$_2$, such as an anomalous linear polarization dependence and changes of the exciton/trion energies.[247] Via time-resolved optical spectroscopy in combination with DFT calculations, Beer *et al*. demonstrated that MoSe$_2$/CrSBr heterostructure has a type-III band alignment and the time-reversal symmetry of its MoSe$_2$ ML is broken due to the magnetic proximity effect (Figure 11F).[248] Experimentally, the proximity-induced spin splitting in this heterostructure is about 1.6 meV.

The out-of-plane magnetization induced valley splitting in heterostructures of ML valley semiconductors and 2D vdW magnetic materials is also explored in theoretical studies. Through constructing the heterostructures of two FM insulators (i.e., CrI$_3$ and CrBr$_3$) and five transition-metal dichalcogenide MLs (i.e., MoS$_2$, MoSe$_2$, MoTe$_2$, WS$_2$, and WSe$_2$), Zhang *et al*. found using DFT calculations and the $\bm{k}\cdot\bm{p}$ model analyses that MoTe$_2$/CrBr$_3$ heterostructure hosts a large valley splitting of about 28.7 meV.[244] Strikingly, they found that WSe$_2$/CrBr$_3$ is a valley-polarized QAHE system with Chern number $C_N = -1$ at the $K$ point while $C_N = 0$ at the $K'$ point (Figure 11G). Based on the DFT calculation studies of the transition-metal dichalcogenide MLs and 2D magnetic materials Ni$Y_2$ ($Y$ = Cl, Br, I), Li *et al*. argued that large valley splitting can be obtained only when the vdW heterostructure has a type-III, instead of a type-I or type-II, band alignment.[249] For the VI$_3$/WSe$_2$ heterostructure, it is demonstrated that its valley splitting can be increased from 1.8 meV at 4% compressive strain to 3.1 meV at 4% tensile strain.[250] A DFT calculation of heterostructures of Cr$_2$Ge$_2$Te$_6$ and 2$H$-M$X_2$ (M = Mo, W, and X = S, Se, Te) MLs suggested that Cr$_2$Ge$_2$Te$_6$/2$H$-M$X_2$ exhibits a stacking dependent valley splitting.[251] Particularly, a giant valley splitting of about 38 meV can be achieved in theoretically designed heterostructure of transition-metal dichalcogenide MoTe$_2$ ML and vdW Janus FM insulator CrSBr ML.[252]

As a result of the valley splitting, there can exist valley-dependent quantum states in heterostructures of ML valley semiconductor and 2D FM insulators. DFT calculations suggested that, the quantum valley Hall effect with a valley Chern number, $C_v = 2$, can show up in the heterostructure of germanene (Ge) and VBi$_2$Te$_4$ (Figure 11H) no matter this heterostructure has either an out-of-plane or in-plane magnetic moment.[245] For MnBi$_2$Te$_4$/Ge heterostructure, it displays QAHE when MnBi$_2$Te$_4$ has an out-of-plane magnetic moment.[245] However, MnBi$_2$Te$_4$/Ge heterostructure has quantum valley Hall effect with a valley Chern number, $C_v = 2$, when the magnetic moment is along the in-plane direction. For NiBi$_2$Te$_4$/Ge heterostructure, it exhibits QAHE and the coexistence of quantum valley Hall effect and QAHE when its magnetic moment is along the out-of-plane and in-plane directions, respectively.[245] Hence, valley-dependent quantum states are closely correlated to the magnetic anisotropy determined magnetic orientation of 2D FM materials in their vdW heterostructures with valley semiconductors.

At the interfaces of 2D vdW magnetic heterostructures, the inversion symmetry is broken, and a strong DM interaction can be produced. As a result of the interplay between an out-of-plane magnetic anisotropy, exchange interaction and DM interaction, 2D vdW magnetic heterostructure are promising platforms to explore the formation of magnetic skyrmions. Employing Lorentz transmission electron microscopy, Wu *et al.* observed Néel-type magnetic skyrmions in WTe$_2$/Fe$_3$GeTe$_2$ heterostructure (Figure 12A).[253] Through growing Fe$_3$GeTe$_2$ under ambient condition, Park and Peng *et al.* fabricated a unique heterostructure of a Fe$_3$GeTe$_2$ layer sandwiched by two oxidized Fe$_3$GeTe$_2$ layers.[254, 255] Similar to WTe$_2$/Fe$_3$GeTe$_2$ heterostructure, they observed Néel-type magnetic skyrmions in this heterostructure. In a heterostructure of Fe$_3$GeTe$_2$ and graphene, Srivastava *et al.* showed the emergence of Rashba-effect induced strong DM interactions at the graphene/Fe$_3$GeTe$_2$ interface.[256] Based on the measured topological Hall effect and DFT calculated magnetic parameters, they argued that the enhanced micromagnetic DM interactions prevails over the MAE, resulting in the formation of magnetic skyrmions in graphene/Fe$_3$GeTe$_2$.

A vdW heterostructure of two 2D FM films, Cr$_2$Ge$_2$Te$_6$ with a weak out-of-plane magnetic anisotropy and Fe$_3$GeTe$_2$ with a strong out-of-plane magnetic anisotropy, is experimentally reported to host two groups of magnetic skyrmions.[257] Below the $T_C$ (~65 K) of Cr$_2$Ge$_2$Te$_6$, the peak and dip features of Hall resistivity measurements showed the formation of magnetic skyrmions on Cr$_2$Ge$_2$Te$_6$ and Fe$_3$GeTe$_2$ sides with opposite polarity in Fe$_3$GeTe$_2$/Cr$_2$Ge$_2$Te$_6$ heterostructure (Figure 12B). However, above 60 K, magnetic skyrmions only exist on the Fe$_3$GeTe$_2$ side. Given that the magnetic skyrmion size (about 100 nm) in Fe$_3$GeTe$_2$/Cr$_2$Ge$_2$Te$_6$ heterostructure is larger than the magnetic domain size in commercial hard disk drives, electric gating is suggested to tune the magnetic skyrmion size, because the magnetic anisotropies of Fe$_3$GeTe$_2$ and Cr$_2$Ge$_2$Te$_6$ can be modified.

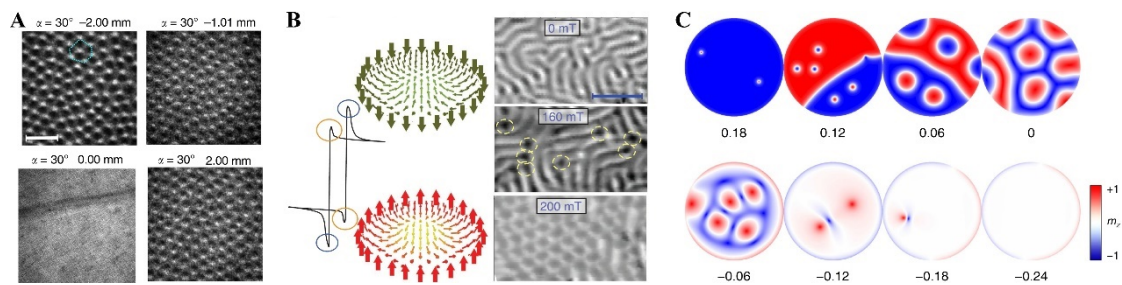

*Figure 12.* A) A Lorentz transmission electron microscopy observation of magnetic skyrmion lattice from under focus to over focus on $WTe_2/Fe_3GeTe_2$ heterostructure at 180 K with a magnetic field of 510 Oe. Reproduced with permission.[253] Copyright 2020, The Author(s). B) Left panel: two sets of topological Hall effect signals observed in $Fe_3GeTe_2/Cr_2Ge_2Te_6$ for temperatures lower than 60 K. The blue and orange circles signify the topological Hall effect on the $Fe_3GeTe_2$ and $Cr_2Ge_2Te_6$ sides, respectively. Skyrmion lattice observed on the $Fe_3GeTe_2$ and $Cr_2Ge_2Te_6$ sides with a magnetic field of 200 mT and a temperature of 20 K. Scale bar: 1 μm. Reproduced with permission.[257] Copyright 2022, The Authors. C) Evolutions of spin textures under the modulations of magnetic anisotropy in $LaCl/In_2Se_3$ heterostructure with a polarization of $In_2Se_3$ along the +z axis. Reproduced with permission.[258] Copyright 2020, The Author(s).

The interface of 2D vdW magnetic heterostructure also gives a window to tune the magnetic anisotropy, which provides a possibility of manipulating magnetic skyrmions. In a vdW magnetic heterostructure of a 2D FM layer LaCl and a 2D ferroelectric layer $In_2Se_3$, it is theoretically shown that magnetic skyrmions in the form of bimerons can be created and annihilated in LaCl layer by switching the ferroelectric polarization of $In_2Se_3$ layer.[258] The core mechanism for such control of the magnetic skyrmions in $LaCl/In_2Se_3$ heterostructure is the polarization reversal induced magnetic anisotropy switching (Figure 12C). As a result of the controllable magnetic anisotropy of 2D Janus magnetic CrSeI, which is influenced by the polarization of 2D ferroelectric layer $In_2Te_3$, a switching between intrinsic magnetic skyrmion and high-temperature FM state is put forward in a vdW magnetic heterostructure $CrSeI/In_2Te_3$ by reversing the ferroelectric polarization of $In_2Te_3$.[259] In a vdW magnetic heterostructure of the FM $CrCl_3$ and 2D topological insulator $WTe_2$ ML, a combination of DFT calculations and micromagnetic simulations showed an out-of-plane magnetic field medicated Néel-type skyrmions-bimeron-ferromagnet phase transition, which is largely driven by the competition between magnetic dipole-dipole interactions and an out-of-plane MCA.[260] Similarly, a magnetic field-controlled transition cycle between a Néel-type magnetic skyrmion lattice and FM states are theoretically demonstrated in $CrTe_2/WTe_2$ heterostructure.[261] Through DFT calculations and atomistic spin simulations, Li *et al*. predicted that a compressive strain leads to stabilizing zero-field skyrmions with diameters close to 10 nm in a vdW heterostructure Fe$_3$GeTe$_2$/germanene.[262] The origin of these unique skyrmions can be attributed to the high tunability of DM interaction and MCA energy by strain.

**5. Methods to Tune Magnetic Anisotropy of 2D vdW Magnets**

Since the magnetic anisotropy of 2D vdW magnetic materials is closely linked to their electronic band states near the Fermi level, it can be effectively modulated by approaches that alter either the band structure or the Fermi level position. This review highlights several widely adopted strategies for tuning the properties of vdW magnetic materials, such as alloying, electrostatic gating, chemical doping, strain engineering, and interfacial engineering.

Because the SOC of ligand atoms is of importance to magnetic anisotropy of 2D vdW magnetic materials, altering this SOC is expected to change magnetic anisotropy. In view of this, alloying of ligand atoms is adopted to tune the magnetic anisotropy in vdW magnetic materials. The alloyed CrCl$_{3-x}$Br$_x$ is taken as an example here. As shown in Figure 13A, the subset of molecular orbital levels relevant to its charge-transfer processes is dependent on the ligand Cl or Br atoms due to the energy difference between 4$p$ orbitals of Cl or 5$p$ orbitals of Br. As a result, the effective SOC in CrCl$_{3-x}$Br$_x$ depends on the mixing ration between Cl and Br who have a different SOC. Considering this, Abramchuk *et al*. grew CrCl$_{3-x}$Br$_x$ to explore the control of its magnetic anisotropy (Figure 13B).[263] The CrCl$_{3-x}$Br$_x$ samples with different Br contents ($x$) exhibit different colors (Figure 13C). By studying the difference, $\Delta M = M(H_\parallel) - M(H_\perp)$, as a function of magnetic field (the upper panel in Figure 13D), they found that CrCl$_{3-x}$Br$_x$ samples with $x < 2$ exhibit a positive $\Delta M$ in the positive field direction, reminiscent of in-plane easy-axis. This behavior is inversed in samples with x > 2 where the easy axis acquires a dominant out-of-plane component. By defining an empirical anisotropy factor $\alpha = (H_h - H_e)/H_e$ where $H_h$ and $H_e$ mark the onset of saturation in magnetization curves along the hard and easy axes, they found that the magnetic anisotropy of CrCl$_{3-x}$Br$_x$ crosses the boundary between in-plane and out-of-plane at $x = 2$ (the bottom panel in Figure 13D). Considering the different SOC of halogen elements, it is expected that the magnetic anisotropy Cr$X_3$ ($X$ = Cl, Br, I) can be modulated by alloying Cr, Br and I.

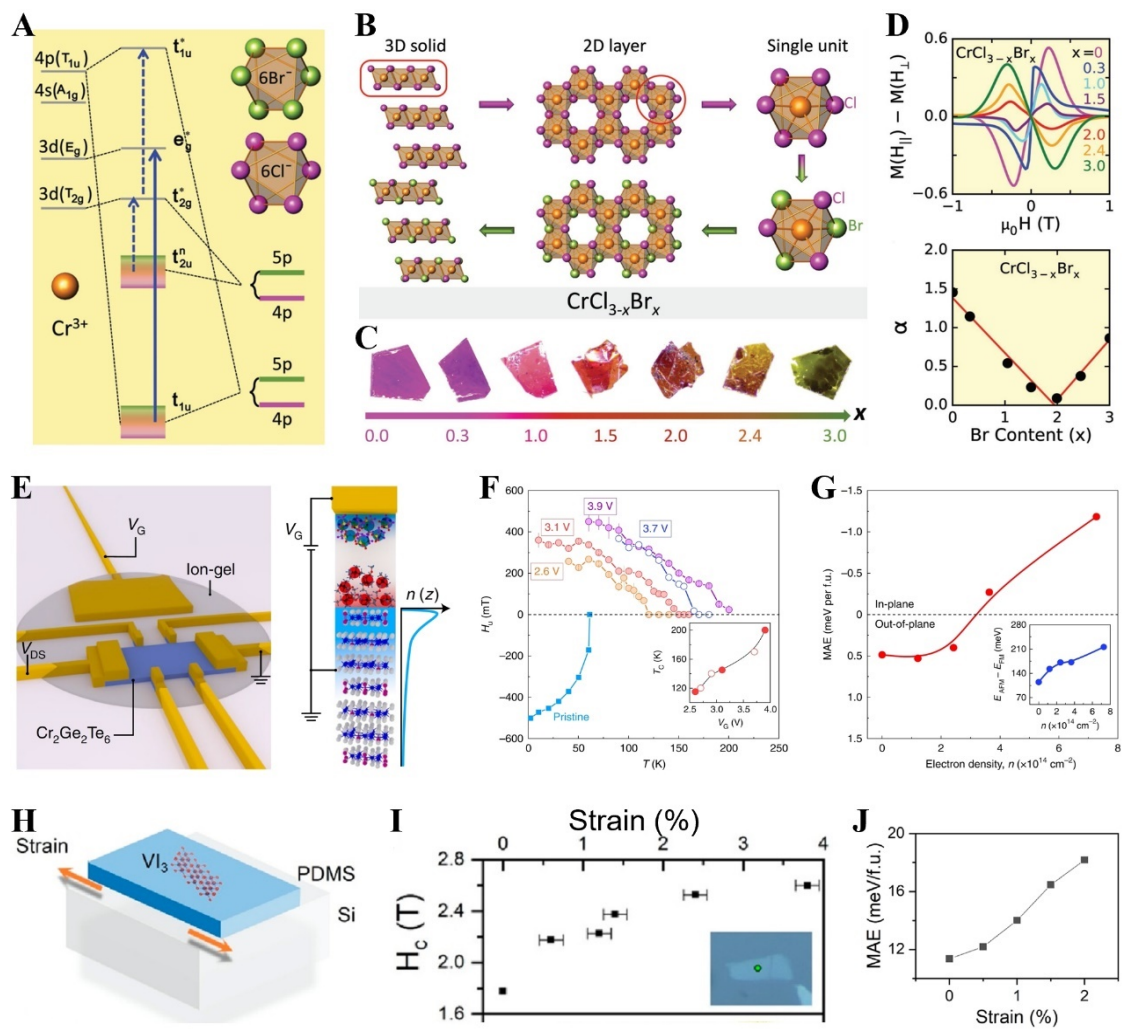

*Figure 13.* A) A subset of molecular orbital levels relevant to charge-transfer processes in $CrCl_{3-x}Br_x$. Arrows show the allowed transitions between the levels. B) Concept of the mixed halide chemistry where Br replaces Cl within each honeycomb layer in $CrCl_{3-x}Br_x$. C) Optical images of $CrCl_{3-x}Br_x$ with different x. D) Upper panel: the difference $\Delta M$ as a function of magnetic field for $CrCl_{3-x}Br_x$. Bottom panel: the anisotropy factor $\alpha$ (defined in the text) plotted as a function of Br content (x) in $CrCl_{3-x}Br_x$. A), B), C) and D) are reproduced with permission.[263] Copyright 2081, WILEY-VCH Verlag GmbH & Co. KGaA, Weinheim. E) Left panel: schematic of an EDLT device based on $Cr_2Ge_2Te_6$ thin film. Right panel: the carrier density distribution, n, along the thickness direction, z. F) $H_u$ as a function of gate bias. Squares represent values measured for a pristine bulk sample $Cr_2Ge_2Te_6$, while filled and open circles represent data from EDLT devices 1 and 2, respectively. Inset: the dependence of $T_C$ on gate bias ($V_G$). G) Calculated MAE as a function of electron density, n, in $Cr_2Ge_2Te_6$. The change of $E_{AFM} - E_{FM}$ at different doping densities is shown in the inset. Solid lines are guides to eyes. Reproduced with permission.[10] Copyright 2020, The Author(s), under exclusive licence to Springer Nature Limited. H) Schematic diagram of applying an in-plane

*tensile strain in the VI$_3$ flake. I) The extracted averaged coercivity (H$_C$) as a function of applied tensile strain at 10 K in the strained VI$_3$ flake. J) DFT calculated strain-dependent MAE in VI$_3$ ML. H), I) and J) are reproduced with permission.*[264] *Copyright 2022, American Chemical Society.*

Alloying magnetic ions is also put forward to engineer magnetic anisotropy in vdW magnetic materials. Experimentally, motivated by the different magnetic anisotropies in the isostructural FePS$_3$ (i.e. Ising-like) and NiPS$_3$ (i.e., XXZ-like), Selter and Lee *et al.* studied the magnetic properties of Ni$_{1-x}$Fe$_x$PS$_3$.[265, 266] They found that Ni$_{1-x}$Fe$_x$PS$_3$ shows a turnover from the XXZ-like to Ising-like magnetic anisotropy through x ≈ 0.1. Given the weak SOC of Cr$^{3+}$ ions in CrCl$_3$ ML, alloying Cr$^{3+}$ magnetic ions with isovalent Mo$^{3+}$ or W$^{3+}$ ions is theoretically proposed to strengthen its magnetic anisotropy,[144] because Mo$^{3+}$ or W$^{3+}$ ions have the same electronic configuration as Cr$^{3+}$ magnetic ions but much larger SOC. DFT calculations showed that the doped CrMoCl$_6$ and CrWCl$_6$ have an out-of-plane magnetic anisotropies,[144] which is different from the in-plane magnetic anisotropy in the pristine CrCl$_3$ ML.

Electrostatic gating can tune the magnetic anisotropy of vdW magnetic materials by introducing electron doping. In electron-doped (~10$^{14}$ cm$^{-2}$) Cr$_2$Ge$_2$Te$_6$ thin film in an electric double-layer transistor (EDLT) devices (Figure 13E), Verzhbitskiy *et al.* revealed via angle-dependent magnetoresistance measurements that the magnetic easy axis for heavily doped Cr$_2$Ge$_2$Te$_6$ is along the in-plane direction (Figure 13F), in contrast to the out-of-plane easy axis of undoped Cr$_2$Ge$_2$Te$_6$. Their DFT calculations indicated that the magnetic anisotropy of electron doped Cr$_2$Ge$_2$Te$_6$ is out-of-plane for a low electron density but become in-plane for a high electron density (Figure 13G), which is consistent in general trend with their experimental observation of magnetic easy axis switching in the heavy doping limit.

Electrochemical intercalation of organic ions is experimentally demonstrated to be an effective and convenient way to tune the magnetic anisotropy of Cr$_2$Ge$_2$Te$_6$. Wang *et al.* reported that a hybrid superlattice (TBA)Cr$_2$Ge$_2$Te$_6$ is formed when Cr$_2$Ge$_2$Te$_6$ is intercalated with tetrabutyl ammonium cations (TBA$^+$).[267] Different from the out-of-plane magnetic easy axis of Cr$_2$Ge$_2$Te$_6$, the magnetic easy axis of (TBA)Cr$_2$Ge$_2$Te$_6$ lies in the *ab*-plane. This means a TBA$^+$-intercalation induced magnetic easy axis switching in Cr$_2$Ge$_2$Te$_6$. Their DFT calculations revealed that (TBA)Cr$_2$Ge$_2$Te$_6$ is electron doped with conducting states coming from the $d_{xz}$ and $d_{yz}$ orbitals. Their calculations further indicated that the doublet $d_{xz}$ and $d_{yz}$ orbitals combining with other 3d orbitals would contribute an effective nonzero orbital magnetic moment M$_L$ // c = 0.0058 $\mu_B$ and M$_L$ // a = 0.0281 $\mu_B$, mainly lying on the *ab*-plane. Due to the SOC of Cr ions, the planar orbital magnetic moments favor the spin alignment within the *ab*-plane in $Cr_2Ge_2Te_6$ intercalated with tetrabutylammonium cations.

One feasible route to modify the magnetic anisotropy of vdW magnetic materials is the application of hydrostatic or tensile pressure. By applying hydrostatic pressure up to 2 GPa to $Cr_2Ge_2Te_6$, Lin *et al*. reported that their anisotropic magnetoresistance measurements indicated that its magnetic anisotropy is switched from out-of-plane to in-plane when the hydrostatic pressure is larger than 1.0 GPa.[268] Their DFT calculations confirmed the giant MAE change with moderate pressure and assigned its origin to the increased off-site SOC of Te due to a shorter Cr-Te distance. However, a separate FMR study found that the magnetic orientation in $Cr_2Ge_2Te_6$ remains to be out-of-plane up to 2.39 GPa, even though MAE was found to decrease under hydrostatic pressure.[269] The pressure induced reductions of out-of-plane MAE were also reported for two other vdW magnetic materials, $CrI_3$ and $CrBr_3$.[270, 271]

A strain tunability of the out-of-plane magnetic anisotropy is reported in $VI_3$.[264] By employing thin polydimethylsiloxane (PDMS) as a stretcher (Figure 13H), Zhang *et al*. carefully investigated the effect of tensile strain on the magnetic anisotropies of $VI_3$ flakes. They performed *in situ* magnetic circular dichroism measurements at 10 K in a reflection geometry of $VI_3$ flakes with an external magnetic field along the *c*-axis of the sample. They found that the FM ground state of $VI_3$ flakes is preserved, and the coercive force gradually increases from 1.78 T in the initial state to a maximum of 2.6 T under a 3.8% tensile strain, as the tensile strain increases (Figure 13I). Their DFT calculations suggested that the increased MAE under tensile strain contributes to the enhancement of coercivity (Figure 13J). Meanwhile, they found the strain tunability on the coercivity of $CrI_3$, with a similar crystal structure to $VI_3$, is negligible in comparison with that in $VI_3$ under the same tensile strain. They argued that the different strain tunability of magnetic anisotropy between $VI_3$ and $CrI_3$ may be mainly related to the strong MAE induced by unquenched orbital moment in $VI_3$ but by ligand SOC in $CrI_3$. The strain tunability of magnetic anisotropy in $VI_3$ highlights its potential for integration into 2D vdW heterostructures.

Theoretically, it is also shown that strain can tune the magnetic anisotropy of 2D vdW magnetic materials. Webster *et al*. performed a DFT study on the MAE of $CrX_3$ (X = Cl, Br, and I) ML under the influence of a biaxial strain. They found that the MAE increases when a compressive strain is applied in $CrI_3$ while an opposite trend is observed in $CrBr_3$ and $CrCl_3$.[143] In particular, the MAE of $CrI_3$ can be increased by 47% with a compressive strain of 5%. For the $Cr_2Ge_2Te_6$ bilayer, a DFT study suggested that the direction of its magnetic easy axis can be converted from out-of-plane to in-plane due to the increase of a compressive strain.[272] Using DFT calculations, Kim *et al*. demonstrated that the out-of-plane MCA of $Fe_3GeTe_2$ ML and bilayer is retained with little change in the magnetic moments for the tensile strain.[198] On the other hand, abrupt decreases in their $E_{MCA}$ and changes in the magnetic moments are prominent for the compressive strain. In particular, $Fe_3GeTe_2$ bilayer even exhibits a sign change of $E_{MCA}$, transforming from out-of-plane to in-plane magnetic anisotropy. In $Fe_5GeTe_2$ ML, a compressive strain induced magnetic anisotropy switching from in-plane to out-of-plane is shown by DFT calculations.[208]

It has been shown that hole- or electron-doping can modulate the magnetic anisotropy of 2D vdW ferromagnets. By changing Fe content (i.e. hole doping), $Fe_{3-x}GeTe_2$ exhibits a significant lower $T_C$ and largely reduced magnetic coercivity than $Fe_3GeTe_2$.[196] The large reduction in magnetic coercivity implies that $Fe_{3-x}GeTe_2$ has significantly smaller magnetic anisotropy than $Fe_3GeTe_2$.[196] The doping dependent changes in the electronic structures unveiled by XMCD, ARPES and DFT calculations showed that the decreased out-of-plane magnetic anisotropy is due to the reduction of the electron pockets which possess strong SOC induced band splitting that dominantly contributes to the out-of-plane magnetic anisotropy (Figure 14A).[196] It is noteworthy that $Fe_3GaTe_2$, which is isostructural to $Fe_3GeTe_2$, also exhibits doping-driven tunable magnetic anisotropy.[273] This finding shows the intricate connection between electronic structures and magnetic anisotropies in 2D vdW magnetic materials. For $CrBr_3$ ML, a DFT study showed that hole doping with an experimentally attainable hole density can boost its magnetic anisotropy as well as effective FM exchanges, thus leading to the room-temperature ferromagnetism.[274] Given the dependence of MAE on the electron or hole doping in $CrI_3$ ML, DFT studies indicated doping electron into $CrI_3$ ML via chemical adsorption of alkali and alkaline-earth elements can tune its magnetic anisotropy from out-of-plane to in-plane.[135, 275]

Interface engineering is another effective approach to tune the magnetic anisotropy of 2D vdW magnetic materials and their heterostructures. By growing heterostructures of $Fe_3GeTe_2$ or oxidized $Fe_3GeTe_2$ with different thicknesses (Figure 14B), Kim *et al*. found a nonmonotonic behavior of an out-of-plane magnetic anisotropic field which originates from the competition between the interface and bulk MCA.[276] This is understandable because the interface MCA originating from only a few ML adjacent to the camping layer decreases with increasing $Fe_3GeTe_2$ thickness whereas the bulk one shows the opposite behavior. They even achieved local modulation of the out-of-plane magnetic anisotropy in the $Fe_3GeTe_2/WSe_2$ heterostructure, attributing it to the presence of the adjacent ultrathin $WSe_2$ layer, which possesses strong SOC. Interestingly, Idzuchi *et al*. reported that the out-of-plane magnetic anisotropy of $Cr_2Ge_2Te_6$ thin flakes is notably enhanced by using its heterostructure interface with an AFM insulator NiO.[277] They attributed this enhancement to interfacial strain and the bonding interactions between the 3d electrons of the transition metal and the 2p electrons of oxygen.

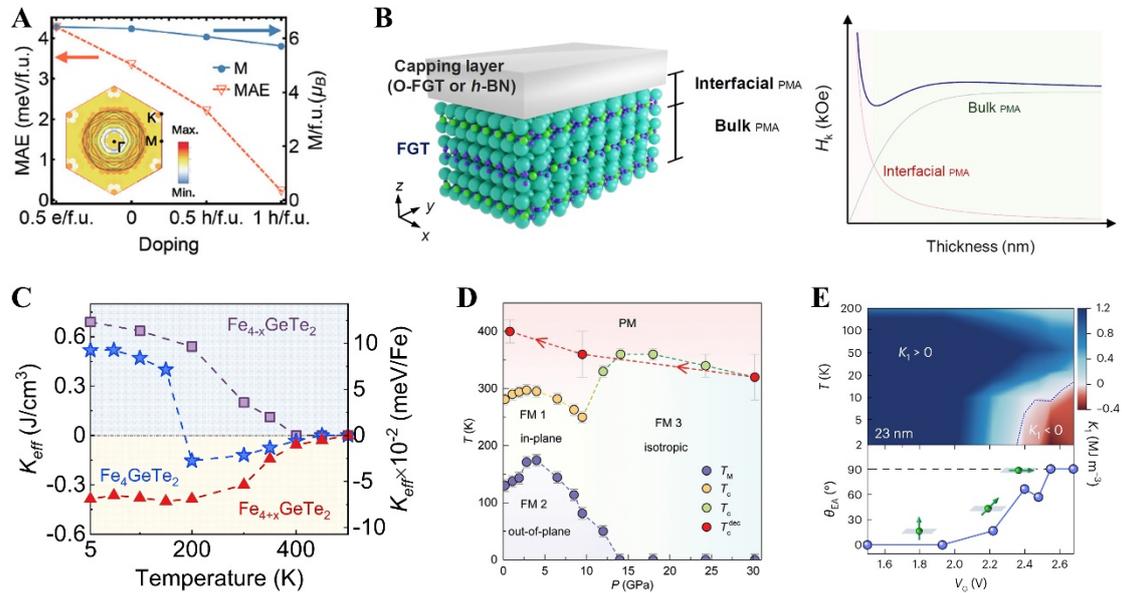

*Figure 14. A summary of the tuned magnetic anisotropies of $Fe_nGeTe_2$ (n = 3, 4, and 5) by different methods. A) MAE and magnetization per formula unit (f.u.) as a function of doping in $Fe_3GeTe_2$. The insert illustrates the momentum-resolved MAE for one hole/f.u. doped $Fe_3GeTe_2$. Reproduced with permission.[196] Copyright 2020, American Chemical Society. B) Left panel: schematic of the range from bulk and interfacial out-of-plane magnetic anisotropies in an $Fe_3GeTe_2$-based heterostructure. Right panel: the competition between bulk and interfacial out-of-plane magnetic anisotropies depending on the thickness of $Fe_3GeTe_2$. Reproduced with permission.[276] Copyright 2021, American Chemical Society. C) Effective magnetic anisotropies, $K_{eff}$, as a function of temperature in $Fe_{4-x}GeTe_2$, $Fe_4GeTe_2$, and $Fe_{4+x}GeTe_2$. Reproduced with permission.[278] Copyright 2023, The Author(s). D) Temperature-pressure phase diagram of $Fe_5GeTe_2$. Reproduced with permission.[279] Copyright 2022, Wiley-VCH GmbH. E) The color mapping for the temperature-dependent first-order constant, $K_1$, for magnetic anisotropy of $Fe_5GeTe_2$ with a thickness of 23 nm under different gate voltages ($V_G$). The green arrows together with a plane show the direction of the magnetic easy axes. Reproduced with permission.[280] Copyright 2022, The Author(s), under exclusive license to Springer Nature Limited.*

Similar to $Fe_3GeTe_2$, the strong out-of-plane magnetic anisotropy of $Fe_3GaTe_2$ can be regulated using the nondestructive vdW interfacial magnetochemistry. By growing vdW heterostructures composed of $Fe_3GaTe_2$ and nonmagnetic vdW materials $MoS_2$,

WSe$_2$, or Bi$_{1.5}$Sb$_{0.5}$Te$_{1.7}$Se$_{1.3}$, the out-of-plane magnetic anisotropy of Fe$_3$GaTe$_2$ at the room temperature is reported to be weakened by 26.8%, though the $T_C$ is enhanced up to 400 K.[202] The analysis of the atom-resolved SOC energy based on DFT calculations further reveals that the interfacial SIA in these heterostructures mainly originates from the contributions of Fe and Te atoms in Fe$_3$GaTe$_2$ layer.[202] Due to the interlayer proximity effect, the total contribution from Fe and Te atoms to SIA in the Fe$_3$GaTe$_2$ layer decreases from 3.13 to 1.89 meV/atom, ultimately leading to the experimentally observed weakening of the magnetic anisotropy in these heterostructures.

The vdW heterostructures composed of 2D metallic and insulating ferromagnets also provide interesting platforms to explore the effect of the interface coupling on magnetic anisotropy. By constructing vdW heterostructure of metallic Fe$_3$GeTe$_2$ and insulating VI$_3$, Tang *et al*. demonstrated via anomalous Hall effect measurement that the coercive field of Fe$_3$GeTe$_2$ is significantly increased by ~150%.[281] Their DFT calculations showed that Fe$_3$GeTe$_2$/VI$_3$ heterostructure has an increased out-of-plane MAE than Fe$_3$GeTe$_2$, which is attributed to the spin coupling of the interface between Fe$_3$GeTe$_2$ and VI$_3$.

Interestingly, Wang *et al*. found that the magnetic anisotropy of Fe$_4$GeTe$_2$ can be precisely controlled along either out-of-plane or in-plane through tailoring its Fe concentration.[278] As shown Figure 14C, although the magnetic anisotropy in Fe$_4$GeTe$_2$ is temperature dependent, Fe$_{4-x}$GeTe$_2$ or Fe$_{4+x}$GeTe$_2$ always present a definite direction of magnetic anisotropy at all temperatures. It is notable that, when the Fe concentration decreases from Fe$_{4+x}$GeTe$_2$ to Fe$_{4-x}$GeTe$_2$, the magnetic anisotropy turns from in-plane to out-of-plane. Such varied magnetic anisotropy may originate from the out-of-plane MCA variation during the stoichiometry change.

As a powerful way to modify the electronic structure near the Fermi level of materials, the high-pressure technique is employed to control the magnetic anisotropy of Fe-based vdW magnetic materials. By analyzing the anomalous Hall effect, magnetoresistance, and high-pressure X-ray diffraction, Li *et al*.[279] observed a pressure-driven magnetic phase transition among three FM phases with different magnetic easy axes, including the FM1 phase with an in-plane magnetic anisotropy, FM2 phase with an out-of-plane magnetic anisotropy, and FM3 phase with nearly isotropic magnetization (Figure 14D). Based on Hall measurements, they found that the dominant carriers of Fe$_5$GeTe$_2$ near the Fermi surface for the Hall effect are pressure-dependent. This indicates the pressure drives a Fermi surface reconstruction and changes the band structure in Fe$_5$GeTe$_2$, and finally leads to an effective control of its magnetic anisotropy. In Fe$_3$GaTe$_2$, a similar continuously tunable magnetic anisotropy from out-of-plane to in-plane is achieved via the application of pressure, which is also ascribed to the pressure-driven Fermi surface reconstruction.[282]

Using electrical gating, Tang *et al.* reported the gate-tunable modulation of magnetic anisotropy, from an initial out-of-plane to a canted orientation and then finally to an in-plane orientation, in vdW ferromagnet $Fe_5GeTe_2$ (Figure 14E).[280] They also showed that the magnetic easy axis of $Fe_5GeTe_2$ continuously rotates via a spin-flop pathway and can be modulated in a range from 2.11 to -0.38 MJm$^{-3}$. Physically, the cations (such as Li$^+$ cations) of gating electrolytes are electrically driven into the vdW gaps of $Fe_5GeTe_2$ bulk without a structural phase transition or chemical reaction therein and finally cause extreme electron doping in $Fe_5GeTe_2$. As a result of doping, the itinerant electron density in $Fe_5GeTe_2$ can be tuned by gate, which is thus responsible for the transitions between magnetic states with different magnetic easy axes.

Theoretically, many DFT studies show that constructing heterostructures of 2D vdW magnetic materials with other nonmagnetic materials can effectively tune the magnetic anisotropy. By building a multiferroic heterostructure of $Cr_2Ge_2Te_6$ and ferroelectric $In_2Se_3$ ML, Gong *et al.* unveiled that the reversal polarization of $In_2Se_3$ switches the MCA of $Cr_2Ge_2Te_6$ between out-of-plane and in-plane orientations.[283] Their detailed analysis suggested that the interfacial hybridization is the cause of the MCA switching in $Cr_2Ge_2Te_6/In_2Se_3$. Compared with the pristine $Cr_2Ge_2Te_6$ ML, it is shown that the MAE of $Cr_2Ge_2Te_6/PtSe_2$ heterostructure is enhanced by 70%.[284] An analysis of DFT results with a model Hamiltonian revealed that both DM interactions and SIA contribute to the enhancement of the MAE in $Cr_2Ge_2Te_6/PtSe_2$. Through building heterostructures of $CrX_3$ (X = Cl, Br, I) ML and the ferroelectric AlN bilayer, Zhu *et al.* demostrated that the magnetic easy axis of $CrX_3$ ML can be switched between out-of-plane and in-plane directions when the polarization direction of the AlN bilayer is reversed.[285] They also showed that the different effects of the AlN bilayer on the modulation of the magnetic easy axis of $CrX_3$ ML are related with their different interfacial charge transfer/redistribution across the interfaces. A study of heterostructure of $MnBi_2Te_4$ ML and 2D ferroelectric $III_2$-$VI_3$ (III = Al, Ga, In; VI = Se, Te) substrates found polarization reversal induced reorientation of magnetic easy axis.[286] Such reorientation is attributed to the sensitivity of the MCA of $MnBi_2Te_4$ ML to the charge redistribution and interfacial hybridization in $MnBi_2Te_4/III_2$-$VI_3$ heterostructure.

Finally, it is suggested that the magnetic anisotropy of a 2D antiferromagnet can be greatly enhanced via stacking it on a magnetic substrate.[287] Such enhancement arises from a sublattice-dependent interlayer magnetic interaction that strongly couples with the interlayer stacking order and magnetic order of the substrate layer. To demonstrate this enhancement, Xiao *et al*. constructed a vdW magnetic heterostructure of MnPS$_3$ bilayer and CrCl$_3$.[287] Although MnPS$_3$ has a very weak MAE, their DFT calculations suggested that the MAE of MnPS$_3$/CrCl$_3$ is 40 times as large as that of MnPS$_3$. Meanwhile, there is a strong coupling between the MAE and stacking order in MnPS$_3$/CrCl$_3$. Such tunable magnetic anisotropy in vdW heterostructures is of interest for stabilizing and manipulating 2D magnetism.

## 6. Conclusion and Outlook

Throughout this review, we have highlighted the critical role of magnetic anisotropy in stabilizing and shaping long-range magnetic orders in 2D systems, as well as in determining the electronic, magnetic, and topological properties of 2D vdW magnetic materials and their heterostructures. While the well-known Mermin-Wagner theorem[3] states that an isotropic 2D Heisenberg model cannot support long-range magnetic ordering at finite temperatures, experimental evidence from ultrathin magnetic films[30] and atomically thin 2D vdW magnetic MLs clearly shows the existence of 2D magnetism in various systems. This apparent contradiction arises because the Mermin-Wagner theorem does not account for magnetic anisotropy, which is a key symmetry-breaking factor in real 2D magnetic materials. The core mechanism enabling the stabilization of 2D magnetism lies in the ability of magnetic anisotropy to break the continuous rotational symmetry, thereby suppressing the divergence of spin-wave excitations that would otherwise destroy long-range orders at finite temperatures. [53] Moreover, magnetic anisotropy not only stabilizes magnetism but also dictates the preferred direction of magnetizations in 2D vdW magnetic materials and their heterostructures. Since the electronic, magnetic, and topological properties of these materials are often highly sensitive to the magnetization direction, magnetic anisotropy emerges as a central parameter in the design and realization of future 2D spintronic and quantum devices.

Given the importance of magnetic anisotropy in 2D vdW magnetic materials and their heterostructures, we have also reviewed its underlying mechanisms and tuning strategies explored in both theoretical and experimental studies. Among the three primary components of magnetic anisotropy, the MSA is the most straightforward, as it arises solely from the magnetic moments and lattice geometry, as described by Eq. (3). In contrast, MCA and exchange anisotropy are more complex, because they depend intricately on the electronic structure of the specific magnetic material. Both typically originate from either the strong SOC of surrounding ligands or the unquenched orbital magnetic moments of the magnetic ions, or a combination of both. Fundamentally, MCA and exchange anisotropy are governed by the band states near the Fermi level, making them sensitive to the changes in the electronic environment. Based on this understanding, a variety of approaches have been demonstrated to effectively tune magnetic anisotropy, including electrostatic gating, charge doping (via holes or electrons), cation or anion alloying, application of strain or hydrostatic pressure, and engineering of interfaces or heterostructures. These strategies provide a versatile toolbox for tailoring the magnetic properties of 2D vdW magnetic materials, thus enabling their optimization for future spintronic and quantum applications.

Despite the intensive research on 2D vdW magnetic materials and their heterostructures in recent years, the understanding and control of their magnetic anisotropies remain challenging and incomplete. A primary unresolved issue is the lack of theoretical consensus on which anisotropic interaction terms are most critical in determining magnetic anisotropy. While it is broadly accepted that strong ligand SOC and unquenched orbital magnetic moments play essential roles, the exact contributions of various symmetry-allowed anisotropic interactions remain debated. Even in the well-studied systems like $CrI_3$, significant uncertainty persists regarding the magnitudes and relevance of anisotropic exchange interactions, SIA, Kitaev interactions, and second-neighbor DM interactions.[288] Compounding the issue, computational predictions such as the MCA of multilayer $Cr_2Ge_2Te_2$ are found to be highly sensitive to the choice of exchange-correlation functional in DFT calculations.[151] Another critical yet difficult challenge is achieving sufficiently strong magnetic anisotropy, which is vital for stabilizing robust and high-temperature 2D magnetism. Currently, most 2D vdW magnets rely on SOC arising from $p$ or $d$ orbitals of $3d$ elements, which is often not strong enough. In contrast, $5d$ and rare-earth elements have significantly stronger SOC, making them promising candidates for enhancing magnetic anisotropy. Therefore, strategies such as doping $5d$ and rare-earth elements into existing 2D vdW magnetic materials hold substantial potential for realizing stronger anisotropic effects and advancing the field of 2D magnetism.

Last but not least, the magnetic anisotropy of moiré magnetism in twisted 2D vdW magnetic materials remains largely unexplored. Recent studies have shown that moiré engineering in these systems can give rise to unconventional nanoscale magnetism, including the coexistence of FM and AFM domains, as well as non-collinear spin textures in twisted $CrI_3$ bilayers.[289-294] Moreover, magnetic skyrmions have been observed in twisted heterostructures composed of a FM ML interfaced with a layered AFM substrate.[295] These discoveries suggest that twist-induced modifications to the electronic density of states[296] and the intrinsic breaking of inversion symmetry in moiré superlattices[297] could give rise to novel mechanisms of magnetic anisotropy—mechanisms that are currently unpredictable within existing theoretical frameworks.

However, the extreme complexity and twist-angle-dependent sensitivity of moiré systems make the theoretical determination of magnetic anisotropy, especially through DFT calculations, highly challenging. This is particularly true for arbitrary twist angles that lead to incommensurate structures, which are computationally intractable using conventional methods. Given the recent success of machine learning in modeling and predicting properties of magnetic materials,[298, 299] there is a promising opportunity to leverage these techniques for studying the magnetic anisotropy in twisted 2D vdW magnetic materials. Such approaches may offer a viable path forward for uncovering and understanding emergent anisotropic phenomena in moiré magnetism.

In conclusion, magnetic anisotropy has emerged as a central factor in the study of 2D magnetic materials and their heterostructures, playing a crucial role in stabilizing long-range magnetic orders in 2D systems. Both experimental and theoretical investigations have significantly advanced our understanding of its underlying mechanisms, particularly the roles of MCA, MSA, and exchange anisotropy. Moreover, a variety of innovative approaches, ranging from strain engineering and pressure tuning to gating, doping, and heterostructure design, have been demonstrated as effective strategies for modulating magnetic anisotropy. As we continue to explore the potential applications of 2D vdW magnetic materials and their heterostructures in next-generation electronic, spintronic, and topological devices, there lies a wealth of opportunities for deeper investigations. In particular, the accurate determination, comprehensive understanding, and precise engineering of magnetic anisotropy will remain vital frontiers, especially in the context of complex phenomena such as the moiré magnetism and twisted heterostructures. The ongoing synergy between theory, experiment, and emerging tools like machine learning holds great promise for unlocking the full potential of 2D magnetic systems.

**Acknowledgements**

This work was supported by the National Key Research Program of China (Grant No. 2024YFA1408303 and 2022YFA1403301), the National Natural Sciences Foundation of China (Grant No. 12474247 and 92165204). R.W. acknowledges support from the United States Department of Energy, Office of Basic Energy Science (Grant No. DE-FG02- 05ER46237).

**Conflict of Interest**

The authors declare no conflict of interest.

## Keywords

Magnetic anisotropy, 2D ferromagnetism and antiferromagnetism, vdW magnetic materials, heterostructure, spin-orbit coupling